\documentclass[preprint,amsmath,amssymb]{revtex4}

\usepackage{graphicx}

\def\slc#1{\setbox0=\hbox{$#1$}           
    \dimen0=\wd0                                 
    \setbox1=\hbox{/} \dimen1=\wd1               
    \ifdim\dimen0>\dimen1                        
       \rlap{\hbox to \dimen0{\hfil/\hfil}}      
       #1                                        
    \else                                        
       \rlap{\hbox to \dimen1{\hfil$#1$\hfil}}   
       /                                         
    \fi}

\begin{document}

\title{Searches for hyperbolic extra dimensions at the LHC}

\author{Henrik Melb\'eus}
\email{melbeus@kth.se}
\affiliation{Department of Theoretical Physics, School of Engineering Sciences,\\
Royal Institute of Technology (KTH) -- AlbaNova University Center,\\
Roslagstullsbacken 21, 106 91 Stockholm, Sweden}

\author{Tommy Ohlsson}
\email{tommy@theophys.kth.se}
\affiliation{Department of Theoretical Physics, School of Engineering Sciences,\\
Royal Institute of Technology (KTH) -- AlbaNova University Center,\\
Roslagstullsbacken 21, 106 91 Stockholm, Sweden}

\newcommand{\ud}{\mathrm{d}}

\begin{abstract}
We investigate a model of large extra dimensions where the internal space has the geometry of a hyperbolic disc. Compared with the ADD model, this model provides a more satisfactory solution to the hierarchy problem between the electroweak scale and the Planck scale, and it also avoids constraints from astrophysics. In general, a novel feature of this model is that the physical results depend on the position of the brane in the internal space, and in particular, the signal almost disappears completely if the brane is positioned at the center of the disc. Since there is no known analytic form of the Kaluza--Klein spectrum for our choice of geometry, we obtain a spectrum based on a combination of approximations and numerical computations. We study the possible signatures of our model for hadron colliders, especially the LHC, where the most important processes are the production of a graviton together with a hadronic jet or a photon. We find that the signals are similar to those of the ADD model, regarding both qualitative behavior and strength. For the case of hadronic jet production, it is possible to obtain relatively strong signals, while for the case of photon production, this is much more difficult.
\end{abstract}

\maketitle

\section{Introduction}\label{Sec:Introduction}

The Large Hadron Collider (LHC) at CERN near Geneva, Switzerland is about to become operative. The searches at the LHC for new physics beyond the Standard Model (SM) will mainly include the potential discoveries of the Higgs boson, supersymmetry, and extra dimensions. In this paper, we will be interested in the third issue, {\it i.e.}, extra dimensions. Indeed, an observation of extra dimensions would be truly revolutionary and would completely change our view of the Universe.

The idea that spacetime could have more than four dimensions was first proposed by Theodore Kaluza \cite{Kaluza:1921tu} and Oskar Klein \cite{Klein:1926tv} at the beginning of the twentieth century. One of the most interesting features of extra dimensions is that they are not ruled out by experiments, provided only that they are compact and small enough to have avoided detection so far. In the scenario known as large extra dimensions, they could even be macroscopically large.

Large extra dimensions were first proposed by Arkani-Hamed, Dvali, and Dimopoulos (ADD) in 1998, their model being known as the ADD model \cite{ArkaniHamed:1998rs,Antoniadis:1998ig}. The novel feature of this model is the assumption that the SM fields are confined to a so-called brane, which is a four-dimensional manifold residing in the full bulk spacetime. This brane is to be identified with ordinary four-dimensional spacetime. Since the SM fields are not allowed to probe the extra dimensions, experimental constraints on their size are avoided to a large extent. Gravity, on the other hand, carries no SM charges and is allowed to probe the extra dimensions. In principle, the assumption that gravity lives in a higher-dimensional spacetime leads to sizable deviations from Newton's inverse-square law at short
distances. However, because of the weakness of the gravitational force relative to the SM forces, Newton's law has only been tested down to distances of the order of micrometers, and hence, the experimental constraints are still quite weak.

One of the main motivations for the ADD model is that it provides a solution to the so-called hierarchy problem between the electroweak scale $M_{\rm ew} \sim 100 \,\, {\rm GeV}$ and the (reduced) Planck scale $M_{\rm Pl} \sim 10^{18} \,\, {\rm GeV}$. Theoretically, the bare Higgs mass is expected to receive higher-order quantum corrections of the order of $M_{\rm Pl}$. This would mean that extreme fine-tuning of the parameters would be needed in order for the electroweak scale to be as low as $100 \,\, {\rm GeV}$. In fact, the ADD model provides a very elegant solution to this problem. Since gravity really propagates in more than four spacetime dimensions, the Planck scale that we observe through gravitational measurements is an effective scale, valid only for energies lower than the inverse of the radius of the internal space. The Planck scale is related to the true fundamental energy scale for gravity through the volume of the internal space. If this volume is large enough, then the fundamental scale for gravity could actually be as low as the electroweak scale. However, there is a problem related to this solution in the ADD model. While the problem of the hierarchy between the electroweak scale and the fundamental scale for gravity is solved, there is a new large hierarchy between the electroweak scale and the inverse of the radius of the internal space. Thus, the hierarchy problem is only reformulated as the question of why the radius of the internal space is so large compared to the electroweak scale.

However, in the ADD model, the internal space is assumed to be flat and compactified on a torus. Thus, one possible solution to the problem of the hierarchy between the electroweak scale and the radius of the internal space is to drop this assumption and instead consider a different geometry. Therefore, it has been argued
in Ref.~\cite{Kaloper:2000jb} that a compact hyperbolic internal space in particular is a better alternative than the flat geometry of the ADD model. Note that, in some sense, the hyperbolic model is a generalization of the ADD model.

In addition, it should be mentioned that there are other models of extra dimensions that include branes. One of the most important models is the so-called Randall--Sundrum (RS) model \cite{Randall:1999vf,Randall:1999ee}, in which two branes are introduced and the SM fields are confined to one of these branes only. Nevertheless, we will not consider such models further.

In this paper, we investigate large extra dimensions with the internal space being a two-dimensional hyperbolic disc. Especially, we study two plausible signals, {\it i.e.}, the reactions $p + p \to {\rm jet} + G$ and $p + p \to \gamma + G$, where $G$ denotes a Kaluza--Klein (KK) mode of the graviton, that could be measured at the LHC using missing-energy techniques. It should be noted that the hyperbolic disc model, like the ADD model, is only an effective theory, which means that it is a non-renormalizable low-energy approximation of a more fundamental theory that is called the ultraviolet (UV) completion of the effective theory.

The phenomenology of the ADD model has been extensively investigated in the literature \cite{Giudice:1998ck,Han:1998sg,Mirabelli:1998rt,Hewett:1998sn,Nussinov:1998jt,Rizzo:1998fm,Cheung:1999zw,Balazs:1999ge}. In particular, signals of the ADD model that are relevant for the LHC have been studied in Ref.~\cite{Giudice:1998ck}. In Ref.~\cite{Leblond:2001ex}, a model with a spherical internal space has been examined. In addition, a model of RS type, which is similar to ours, was considered in Ref.~\cite{Bauer:2006pf}, in the setting of discretized extra dimensions. Finally, hyperbolic extra dimensions could have interesting implications in cosmology, which have been studied in Refs.~\cite{Starkman:2000dy,Starkman:2001xu}, though we will not discuss this issue further in this paper.

This paper is organized as follows. In Sec.~\ref{Sec:TheHyperbolicDiscModel}, we present the hyperbolic disc model for large extra dimensions and obtain an approximate form for the KK spectrum of the graviton in this model. Then, in Sec.~\ref{Sec:Interactions}, we analyze the interactions between the graviton and the SM fields that are relevant for the plausible signals of the model at the LHC. Next, in Sec.~\ref{Sec:NumericalAnalysisOfSignals}, we give our numerical results for the cross sections of the signals discussed in Sec.~\ref{Sec:Interactions}. Finally, in Sec.~\ref{Sec:SummaryAndConclusions}, we summarize our results and present our conclusions.

\section{The hyperbolic disc model}\label{Sec:TheHyperbolicDiscModel}

\subsection{Hyperbolic extra dimensions}

The model that we consider is similar to the ADD model, with the only exception that the internal space is a two-dimensional hyperbolic disc, which is denoted $H^2$. Hence, the geometry of the higher-dimensional spacetime is a product $M^4 \times H^2$, where $M^4$ denotes four-dimensional Minkowski space. The SM fields are assumed to be confined to a four-dimensional brane, while gravity alone probes the extra dimensions. The metric for the six-dimensional spacetime is
\begin{equation}\label{eq:BGMetric}
	(g_{MN}) = {\rm diag} [1,-1,-1,-1,-1,-v^{-2} \sinh^2(vr)],
\end{equation}where $r \in [0,L]$ and $\varphi \in [0,2\pi)$ are polar coordinates and $v$ is the curvature of the disc. The coordinate system is such that $r$ is the physical radial distance between the origin and a point $(r,\varphi)$. We follow the convention that indices in the full spacetime are written as upper-case Roman letters, $M = 0,1,2,3,5,6$, Minkowski indices are written as lower-case Greek letters, $\mu = 0,1,2,3$, and indices in the internal space are written as lower-case Roman letters, $i = 5,6$. Also, $x$ denotes the coordinates in $M^4$ or in the full higher-dimensional spacetime and $y$ the coordinates in $H^2$. Note that $|\det (g_{MN})| = |\det (g_{ij})|$, which means that there is no ambiguity in using the symbol $|g|$ for both of these quantities. The number of extra dimensions is denoted by $d$.

The most common way to hide the extra dimensions is through compactification of the internal space as a quotient space $H^2/\Gamma$, where $\Gamma$ is a discrete subgroup of the isometry group of the internal space. In this paper, we consider instead an internal space with an explicit boundary. The main motivation for this choice is computational simplicity. For an internal space of hyperbolic geometry, it is not possible to solve for the KK spectrum analytically. Instead, numerical calculations are needed, and these are much simpler in a space with a boundary than in a quotient space. We do not attempt to describe the origin of the boundary, but simply to investigate its possible implications. An important consequence of this choice of geometry is that, in contrast to the ADD model, the physical results depend on the position of the brane in the internal space.

The energy-momentum tensor corresponding to the metric \eqref{eq:BGMetric} is $(T_{MN}) = {\rm diag} (-v^2,v^2,v^2,v^2,0,0)$. As mentioned, we do not attempt to justify this solution of Einstein's equations. For a deeper discussion, see Refs.~\cite{Bauer:2006ti,Bauer:2006pf}.

The most important advantage of a hyperbolic space is that it offers the possibility of a more satisfactory solution to the hierarchy problem than the ADD model does, as is described below. Another important advantage is that astrophysical constraints on the lower bound on the fundamental mass scale, which are particularly important in the two-dimensional ADD model \cite{ArkaniHamed:1998nn}, can be avoided to a large extent. Thus, the model can allow for a low value of this mass scale even in the case of two extra dimensions only. The constraints on the parameter space are described in more detail in Sec.~\ref{Sec:Constraints}.

\begin{figure}[htb]
\centering
\includegraphics[width=0.5\textwidth,clip]{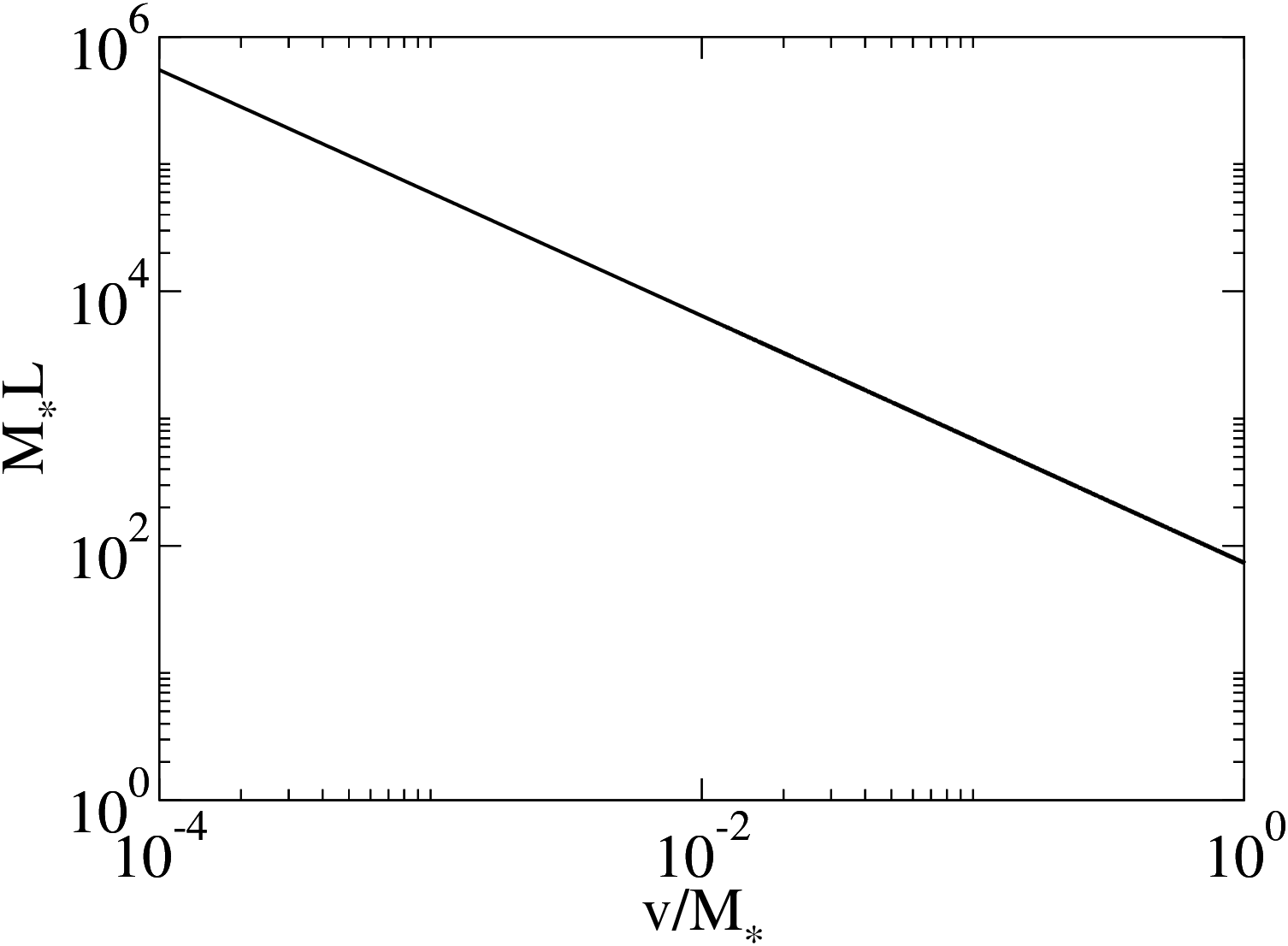}
\caption{The product $M_* L$ as a function of $v/M_*$.}\label{fig:ML}
\end{figure}

Since gravity is the only field probing the internal space, it plays an important part in any phenomenological studies of the model. By assumption, it is governed by the six-dimensional Einstein--Hilbert action
\begin{equation}\label{eq:GravAction}
	S^{\rm (grav)} = M_*^4 \int \!\! \sqrt{|g|} \, \ud^6x \left( R - 2\Lambda \right),
\end{equation}
where $R$ is the Ricci scalar, $\Lambda$ is a cosmological constant, and the mass scale $M_*$ is introduced in order to make the action dimensionless. In the same way as in the ADD model, $M_*$ replaces the Planck scale $M_{\rm Pl}$ as the fundamental mass scale for gravity. The two scales are related through the equation $M_{\rm Pl}^2 = V M_*^4$ \cite{ArkaniHamed:1998nn}, where $V$ is the volume of the internal space. However, note that the definition of the fundamental mass scale differs between authors. In order for the model to provide a solution to the hierarchy problem, we demand that $M_*$ is of the order of $1 \,\, {\rm TeV}$. In the ADD model, where $V_d = (2 \pi L)^d$, this gives the radius $L \sim 10^{31/d} \,\, {\rm TeV}^{-1}$, which is unnaturally large in comparison to $M_*$ if $d$ is not very large. Hence, the hierarchy problem is not really solved, but simply rephrased as the question of why the product $M_* L$ is large. In our model, on the other hand, the volume of the internal space is
\begin{equation}\label{eq:HypDiscArea}
	V = \int \ud V = \int_0^L \!\! \ud r \int_0^{2\pi} \!\! \ud \varphi \sqrt{|g|} = \frac{4\pi}{v^2} \sinh^2 \left( \frac{vL}{2} \right).
\end{equation}
For large $vL$, the volume increases exponentially as a function of the radius. This has the consequence that the relation between the mass scale and the volume of the internal space can be satisfied for $M_* \sim 1 \,\, {\rm TeV}$ without generating a large hierarchy between $M_*$ and $L$, if $v$ is suitably adjusted. This fact is our main motivation for the hyperbolic geometry of the internal space. In Fig.~\ref{fig:ML}, the product $M_* L$ is plotted as a function of $v$. The result is nearly independent of $M_*$ in the range that we are interested in. For $v = M_*$, we obtain the lowest possible value $M_* L \sim 100$, while for smaller $v$, the value of the product is significantly larger. Note that our effective model is supposed to be valid only up to energies of the order of $M_*$, and hence, we do not consider values of $v$ larger than this scale. Thus, the best possible solution to the hierarchy problem in our model is obtained when the curvature $v$ is of the same order of magnitude as the fundamental mass scale. In this case, there is also no new hierarchy problem involving $v$.

\subsection{Kaluza--Klein decomposition of the graviton}

In order to investigate the phenomenology of our model, we now use the ordinary procedure of KK decomposition to reformulate it as an equivalent four-dimensional field theory. In this picture, the graviton field propagating in the full spacetime is represented by an infinite KK tower of particles with different masses, called the KK modes of the graviton. The set of masses is known as the KK spectrum, and each mass corresponds to a quantum of momentum in the internal space.

In the ADD model with $d$ extra dimensions, the graviton living in the full spacetime is expanded in terms of harmonic functions $\exp\left({\rm i} \bar n \cdot \bar y / R\right)$ with corresponding masses $m_{\bar n} = |\bar n| / R$. Here, $\bar n$ is a $d$-dimensional vector with integer entries. In our model, the geometry is more complicated, and this has the result that it is not possible to obtain the KK spectrum analytically. Thus, in this section, we derive approximate expressions for the KK modes and masses in our model.

The starting point is the equations of motion for the graviton. From the action \eqref{eq:GravAction}, it follows that the free equations of motion are the free Einstein equations, {\it i.e.}, $R_{MN} = 0$, where $R_{MN}$ is the Ricci tensor. The dynamics of the graviton is studied by expanding these equations to first order in a perturbation $h_{MN}$ about the background metric \eqref{eq:BGMetric}. The perturbation $h_{MN}$ is interpreted as the massless spin-2 graviton field in the six-dimensional spacetime. We are only interested in the dynamics of the four-dimensional part $h_{\mu\nu}$ of the perturbation, and hence, we make the simplifying approximation of setting all other components to zero, which has often been done in the literature \cite{Leblond:2001ex,Bauer:2006pf,ArkaniHamed:2003vb}. Therefore, the resulting metric is
\begin{equation}
	g_{MN} (x,y) = \left( \begin{array}{cc} \eta_{\mu\nu} + h_{\mu\nu} (x,y) / M_*^2 & 0 \\ 0 & g_{ij} (y) \end{array} \right),
\end{equation}
where $\eta_{\mu\nu}$ is the Minkowski metric and the factor $M_*^{-2}$ ensures that the graviton has the correct dimension, {\it i.e.}, $({\rm mass})^2$. Under the assumption that the unperturbed Einstein equations are satisfied, the detailed derivation of the linearized equations of motion has been performed in Ref.~\cite{Leblond:2001ex}. The resulting equations are
\begin{equation}
	\Delta_{\rm LB} h_{\mu\nu} = 0,
\end{equation}
where $\Delta_{\rm LB} \equiv \nabla^M \nabla_M$ is the Laplace--Beltrami (LB) operator, which is the generalization of the Laplace operator to curved spaces. In terms of a coordinate system $\{x^M\}$,
\begin{equation}
	\Delta_{LB} \psi = \frac{1}{\sqrt{|g|}} \partial_M \left( \sqrt{|g|} g^{MN} \partial_N \psi \right).
\end{equation}
Because of the factorizable geometry, the Laplace--Beltrami operator can be written as $\Delta_{\rm LB} = \square + \Delta_{H^2}$, where $\square \equiv \partial^\mu \partial_\mu$ is the d'Alembert operator in four-dimensional Minkowski space and $\Delta_{H^2}$ is the Laplace--Beltrami operator in the two-dimensional hyperbolic space. In order to reformulate the theory without explicit reference to the extra dimensions, $h_{\mu\nu}$ is expanded in terms of the eigenfunctions of $\Delta_{H^2}$, {\it i.e.},
\begin{equation}\label{eq:KKDecomposition}
	h_{\mu\nu}(x,y) = \sum_n h_{n,\mu\nu}(x) \psi_n(y).
\end{equation}
The coefficient functions $h_{n,\mu\nu}(x)$ are the KK modes of the graviton, satisfying the equations
\begin{equation}\label{eq:KleinGordon}
	\left( \square + m_n^2 \right) h_{n,\mu\nu} = 0,
\end{equation}
where $m_n^2$ is the eigenvalue corresponding to the eigenfunction $\psi_n$. Here, $n$ denotes any general set of Kaluza--Klein indices. From Eq.~\eqref{eq:KleinGordon}, it follows that $m_n$ has the interpretation of the mass of the KK mode $h_{n,\mu\nu}$.

Although no analytic form for the KK spectrum is known, several important results are generally true for the spectrum of the Laplace--Beltrami operator on a Riemannian manifold $M$ (possibly with a boundary). It is assumed that the closure of $M$ is connected and compact, which is true for a hyperbolic disc. The eigenfunctions belong to the Hilbert space $L^2(M)$ of square-integrable functions on $M$, with the inner product given by
\begin{equation}
	\langle f,g \rangle = \int \ud^d y f(y)^* g(y),
\end{equation}
where the star denotes complex conjugation. Then, if the boundary conditions fall into one of the four categories a) Dirichlet conditions, b) Neumann conditions, c) mixed Dirichlet and Neumann conditions, or d) periodic conditions, the following results hold \cite{Chavel:1984}:
\begin{enumerate}
	\item The set of eigenvalues consists of a sequence, $0 = \lambda_1 < \lambda_2 < \ldots \infty$ in the case of Neumann or periodic conditions, or $0 < \lambda_1 < \lambda_2 < \ldots \infty$ in the other cases, and each associated eigenspace is finite-dimensional. \label{item:Point1}
	\item Eigenspaces, belonging to distinct eigenvalues, are orthogonal in $L^2(M)$, which is the direct sum of all eigenspaces.
\end{enumerate}
Thus, when the internal space is compact, the spectrum of the Laplace--Beltrami operator is countable, and it is possible to make a KK expansion of the type \eqref{eq:KKDecomposition}.

Now, we need boundary conditions at $r = 0$ and $r = L$. At $r = 0$, we only have to demand that the solutions are finite. At $r = L$, the possible alternatives are in principle Dirichlet or Neumann conditions, or a combination of both. As mentioned above, the eigenvalues are interpreted as the squared masses of the corresponding KK modes, {\it i.e.}, $\lambda_n = m_n^2$. According to \ref{item:Point1}., there is no massless mode in the spectrum for Dirichlet or mixed conditions, while for Neumann conditions there is such a mode. Hence, in order to obtain the correct low-energy behavior, we impose Neumann conditions at $r = L$.

We also make use of the following result regarding the asymptotic distribution of the eigenvalues, known as Weyl's asymptotic formula: If $N(\lambda)$ denotes the number of eigenvalues in the interval $[0,\lambda]$, counted with multiplicity, then \cite{Weyl:1912}
\begin{equation}
	\lim_{\lambda \to \infty} \frac{N(\lambda)}{\lambda^{d/2}} = \frac{\omega_d V_d}{(2\pi)^d},
\end{equation}
where $d$ is the dimensionality of the manifold $M$, $\omega_d$ is the area of the unit disc in $\mathbb{R}^d$, and $V_d$ is the volume of $M$. Taking $M$ to be the hyperbolic disc, and using Eq.~\eqref{eq:HypDiscArea}, we obtain the result
\begin{equation}
	\lim_{m \to \infty} \frac{N(m^2)}{m^2} = \frac{\sinh^2 \left( vL/2 \right)}{v^2}.
\end{equation}

We now find the general solution of the eigenvalue equation $\Delta_{H^2}\psi = m^2\psi$. These eigenfunctions contain important information on the coupling of the graviton KK modes to SM fields, and they are also the starting point for our numerical investigations of the KK spectrum. For the case of hyperbolic geometry, the Laplace--Beltrami operator is given by
\begin{equation}
	\Delta_{H^2} \psi = - \frac{1}{\sinh (vr)} \frac{\partial}{\partial r} \left[ \sinh (vr) \frac{\partial \psi}{\partial r} \right] - \frac{v^2}{\sinh^2(vr)} \frac{\partial^2 \psi}{\partial \varphi^2}.
\end{equation}
Introducing the dimensionless parameter $\tau \equiv vr$, we obtain the eigenvalue equation
\begin{equation}
	- \frac{1}{\sinh (\tau)} \frac{\partial}{\partial \tau} \left[ \sinh (\tau) \frac{\partial \psi}{\partial \tau} \right] - \frac{1}{\sinh^2(\tau)} \frac{\partial^2 \psi}{\partial \varphi^2} = k^2 \psi,
\end{equation}
where $k^2 \equiv m^2/v^2$. In order to solve this equation, we first expand $\psi$ in the angular direction
\begin{equation}
	\psi (\tau,\varphi) = \sum_{\ell = -\infty}^{\infty} T_\ell (\tau) e^{{\rm i} \ell \varphi},
\end{equation}
and find the radial equation
\begin{equation}\label{eq:EVEqTau}
	- \frac{1}{\sinh (\tau)} \frac{\ud}{\ud \tau} \left[ \sinh (\tau) \frac{\ud T_\ell}{\ud \tau} \right] - \frac{\ell^2}{\sinh^2(\tau)} T_\ell = k^2 T_\ell.
\end{equation}
Introducing	$x \equiv \cosh (\tau)$, we have the equation
\begin{equation}
	\frac{\ud}{\ud x} \left[ (1-x^2) \frac{\ud T_\ell}{\ud x} \right] + \left[ \nu(\nu+1) - \frac{\ell^2}{1-x^2} \right] T_\ell = 0,
\end{equation}
where $\nu(\nu+1) \equiv -k^2$. This is Legendre's associated equation. Its solutions are the associated Legendre functions of the first and second kind \cite{Erdelyi:1953}. Note that these are not the same as the functions encountered {\it e.g.}~in spherical harmonics, since the domain here is $[1,\infty)$ rather than $[-1,1]$. In particular, in this case, $\nu$ is not restricted to integer values in the interval $[-\ell,\ell]$. The associated Legendre functions of the second kind are divergent at $x=1$, and can therefore be discarded as non-physical solutions. The associated Legendre functions of the first kind are the physically acceptable solutions. In what follows, they will be referred to simply as the Legendre functions and they will be denoted by $P_\nu^\ell$. They can be expressed, up to normalization, as
\begin{equation}\label{eq:legendre}
	P_\nu^\ell (x) = \left( \frac{x+1}{x-1} \right)^{\left|\ell\right|/2} F \left(-\nu, \nu+1, 1 + \left|\ell\right|, \frac{1-x}{2}\right),
\end{equation}
where $F$ is the hypergeometric function. It is common to introduce the parametrization $\nu = -1/2 + {\rm i}\rho$, since for real $\rho$, $P_{-1/2 + {\rm i}\rho}^\ell$ is real. Note that, since $\nu = -1/2 \pm {\rm i}\rho$ both give the same value for $\nu(\nu+1)$, we only need to consider values of $\rho$ in {\it e.g.}~the half-plane ${\rm Re}(\rho) \geq 0$. In terms of the parameter $\rho$, the eigenvalues are 
\begin{equation}\label{eq:MRhoEll}
	m_{\rho\ell}^2 = v^2 \left(\frac{1}{4} + \rho^2\right). 
\end{equation}
Finally, the radial eigenfunctions are
\begin{equation}\label{eq:T}
	T_{\rho\ell}(r) = P_{-\frac{1}{2}+{\rm i}\rho}^\ell [\cosh(vr)] = \tanh^{|\ell|} \left( \frac{vr}{2} \right) F \left[ \frac{1}{2} - {\rm i}\rho, \frac{1}{2} + {\rm i}\rho, 1 + |\ell|, -\sinh^2 \left( \frac{vr}{2} \right) \right].
\end{equation}
An important consequence of this result is that, since $F(a,b,c,0) = 1$, $T_{\rho\ell}(0) = 0$ for $\ell \neq 0$. Hence, most of the eigenfunctions are equal to zero at the origin. As is demonstrated in Sec.~\ref{Sec:InteractionLagrangian}, the couplings of the KK modes to SM fields are proportional to the modulus squared of the eigenfunctions, evaluated at the position of the brane. Thus, for the most symmetric location of the brane, at $\tau = 0$, only the $\ell = 0$ modes couple to SM fields. In this case, it is not possible to probe the extra dimensions with the methods that are investigated in this paper, as is discussed in Sec.~\ref{Sec:NumericalAnalysisOfSignals}.

If the parameter $\rho$ is restricted to real values, then the relation \eqref{eq:MRhoEll} implies that the spectrum is restricted to the interval $[v/2,\infty)$. However, there is {\it a priori} nothing to prevent $\rho$ from being complex. The spectrum of the Laplace--Beltrami operator is real and non-negative, and $m = 0$ is in one-to-one correspondence with the constant eigenfunction, but there could possibly exist eigenvalues in the interval $(0,v/2)$, corresponding to values of $\rho$ in the imaginary interval $(0,{\rm i}/2)$. We have numerically investigated the zeros of $\ud T_{\rho\ell}/\ud \tau$ for $\rho$ in this interval and found none. However, we have not been able to prove this result, and such a proof would, of course, be of interest. Nevertheless, we assume in the remainder of this paper that the KK spectrum lies in the interval $[v/2,\infty)$. Thus, under this assumption, there is a mass gap between zero and the mass $m_1 \equiv v/2$ of the first KK mode. The significance of this result is discussed in Sec.~\ref{Sec:Constraints}.

As eigenfunctions, the normalization of the functions $T_{\rho\ell}$ is not determined. It is decided by the normalization of the Lagrangian kinetic terms of the individual KK modes. As the higher-dimensional kinetic terms involving derivatives with respect to $r$ and $\varphi$ become mass terms in the four-dimensional picture, those terms are irrelevant for the following discussion. Since we will later use results for the ADD model from Ref.~\cite{Giudice:1998ck}, we follow their convention, where the relevant kinetic terms are of the forms
\begin{equation}
	\mathcal{L}_{\bar n}^{\rm (kin)} = \frac{1}{2} \partial^\lambda h_{-\bar n}^{\mu\nu} \partial_\lambda h_{\bar n, \mu\nu}.
\end{equation}
Furthermore, since, in the ADD model, $h_{\mu\nu}(x,\bar y) = \sum_{\bar n} h_{{\bar n},\mu\nu}(x) \exp \left( {\rm i} \bar n \cdot \bar y / R \right)$ and $h_{\mu\nu}$ is real, we must have $h_{-\bar n}^{\mu\nu} = h_{\bar n}^{\mu\nu*}$, which means that
\begin{equation}\label{eq:ADDKin}
	\mathcal{L}_{\bar n}^{\rm (kin)} = \frac{1}{2} \partial^\lambda h_{\bar n}^{\mu\nu*} \partial_\lambda h_{\bar n, \mu\nu}.
\end{equation}
In our model, we have
\begin{eqnarray}\label{eq:Normalization}
	\nonumber S^{\rm (kin)} & = & \int \sqrt{|g|} \ud^6 x \frac{1}{2} \partial^\lambda h^{\mu\nu} \partial_\lambda h_{\mu\nu} \\ 
	\nonumber & = & \int \ud^4x \int \sqrt{|g|} \ud^2 y \frac{1}{2} \partial^\lambda \left( \sum_{\rho,\ell} h_{\rho\ell}^{\mu\nu} \psi_{\rho\ell} \right) \partial_\lambda \left( \sum_{\rho^\prime,\ell^\prime} h_{\rho^\prime\ell^\prime,\mu\nu} \psi_{\rho^\prime\ell^\prime} \right) \\
	\nonumber & = & \sum_{\rho,\ell} \sum_{\rho^\prime,\ell^\prime} \int \ud^4 x \frac{1}{2} \partial^\lambda h_{\rho\ell}^{\mu\nu} \partial_\lambda h_{\rho^\prime\ell^\prime,\mu\nu} \int \sqrt{|g|} \ud^2 y \psi_{\rho\ell} \psi_{\rho^\prime\ell^\prime} \\
	\nonumber & = & \sum_{\rho,\ell} \sum_{\rho^\prime,\ell^\prime} \int \ud^4 x \frac{1}{2} \partial^\lambda h_{\rho\ell}^{\mu\nu} \partial_\lambda h_{\rho^\prime\ell^\prime,\mu\nu} \delta_{\rho\rho^\prime} \delta_{-\ell,\ell^\prime} \| \psi_{\rho\ell} \|^2 \\
	\nonumber & = & \sum_{\rho,\ell} \int \ud^4 x \| \psi_{\rho\ell} \|^2 \frac{1}{2} \partial^\lambda h_{\rho,-\ell}^{\mu\nu} \partial_\lambda h_{\rho\ell,\mu\nu} \\ 
	& = & \sum_{\rho,\ell} \int \ud^4 x \| \psi_{\rho\ell} \|^2 \frac{1}{2} \partial^\lambda {h_{\rho\ell}^{\mu\nu}}^* \partial_\lambda h_{\rho\ell,\mu\nu},
\end{eqnarray}
where in the last equality we have used the fact that $\psi_{\rho,-\ell} = \psi_{\rho\ell}^*$, which follows from the results $\left(T_{\rho,-\ell}\right)^* = T_{\rho,-\ell} = T_{\rho\ell}$ and $\exp \left( {\rm i} \ell \varphi \right)^* = \exp \left( -{\rm i} \ell \varphi \right)$. Since $h_{\mu\nu}$ is real, this result implies that $h_{\rho,-\ell}^{\mu\nu} = {h_{\rho\ell}^{\mu\nu}}^*$. The $\delta_{-\ell,\ell^\prime}$, rather than a $\delta_{\ell\ell^\prime}$, in the fourth line comes from the fact that there is no complex conjugation on $\psi_{\rho\ell}$ in the third line. Thus, the normalization of the individual kinetic terms is the same as in Eq.~\eqref{eq:ADDKin}, if we set $\| \psi_{\rho\ell} \|^2 = 1$, which determines the overall normalization of the eigenfunctions.

\subsection{Approximate eigenfunctions}

In principle, all the information about the eigenfunctions is given by Eq.~\eqref{eq:T}. However, in order to better understand their behavior, it is useful to consider a certain approximation of them, which has been adapted from a similar case in Ref.~\cite{Zaldarriaga:1997va}, and is based on the Wentzel--Kramers--Brillouin (WKB) approximation \cite{Shankar:1994}, familiar from quantum mechanics. The approximate expressions for the eigenfunctions given by this approximation also have the advantage that their numerical evaluation requires significantly less computer power than the exact expressions, which has been important for the calculations in Sec.~\ref{Sec:NumericalAnalysisOfSignals}. In order to find these approximate expressions, we introduce the auxiliary functions $u_{\rho\ell}(\tau) \equiv \sqrt{\sinh (\tau)} T_{\rho\ell}(\tau)$. In terms of $u_{\rho\ell}$, Eq.~\eqref{eq:EVEqTau} becomes
\begin{equation}
	- \frac{\ud^2 u_{\rho\ell}}{\ud \tau^2} + \frac{\ell^2-1/4}{\sinh^2 (\tau)} u_{\rho\ell} = \rho^2 u_{\rho\ell}.
\end{equation}
This equation has the form of a one-dimensional Schr{\"o}dinger equation with energy $E = \rho^2$ and potential $V(\tau) = (\ell^2-1/4) / \sinh^2 (\tau)$. Its solutions have differing qualitative behavior depending on the relative magnitudes of $E$ and $V(\tau)$. For $E>V(\tau)$, they are oscillatory, while for $E<V(\tau)$, there is one increasing and one decreasing solution. The turning point $\tau_0$ between the two regions, given by the equation $E = V(\tau_0)$, is $\tau_0 = {\rm arsinh} \left[ \sqrt{(\ell^2-1/4)/\rho^2} \right]$. For $\tau \ll 1$, the solutions are approximately $u_{\rho\ell} (\tau) = \sinh^{\pm|\ell|} (\tau)$. The decreasing solutions diverge at the origin, and correspond to the Legendre functions of the second kind, while the increasing ones correspond to the Legendre functions of the first kind.

Now, for $\tau > \tau_0$, the WKB approximation gives the solutions
\begin{equation}
	u_{\rho\ell} (\tau) = \frac{\sin \left[ \Theta(\tau) \right]}{\left[ \rho^2 - \frac{\ell^2 - 1/4}{\sinh^2 (\tau)} \right]^{1/4}}, \quad \tau > \tau_0,
\end{equation}
where
\begin{equation}
	\Theta(\tau) = \int^\tau \ud \tau^\prime \sqrt{\rho^2 - \frac{\ell^2 - 1/4}{\sinh^2 (\tau^\prime)}} \approx \rho \tau + \varphi_0.
\end{equation}
Here, we have also used the fact that the eigenfunctions are real. For $\tau < \tau_0$, $T_{\rho\ell}$ is small and is approximated as zero, with the understanding that there are no zeros of the derivative in this region. In order for the eigenfunctions to be continuous at $\tau_0$, we set the phase $\varphi_0 = \rho \tau_0$. Thus, the approximate expression for $T_{\rho\ell}$ that we use is given by
\begin{equation}\label{eq:TApprox}
	T_{\rho\ell} (\tau) = \left \{ \begin{array}{ll} \frac{\sin[\rho (\tau - \tau_0)]}{\left[ \rho^2 \sinh^2 (\tau) - (\ell^2 - 1/4) \right]^{1/4}}, & \quad \tau \geq \tau_0 \\
	0, & \quad \tau < \tau_0.
	\end{array}.
	\right.
\end{equation}

The condition for the WKB approximation to hold is
\begin{equation}
	\frac{1}{2\pi} \left| \frac{\ud \lambda}{\ud \tau} \right| \ll 1,
\end{equation}
where
\begin{equation}
	\lambda(\tau) = \frac{2\pi}{[E - V(\tau)]^{1/2}} = \frac{2\pi \sinh (\tau)}{[\rho^2 \sinh^2 (\tau) - (\ell^2 - 1/4)]^{1/2}}.
\end{equation}
This function blows up at $\tau = \tau_0$, where the WKB approximation is generally not valid. 

\begin{figure}[htb]
\centering
\includegraphics[width=\textwidth,clip]{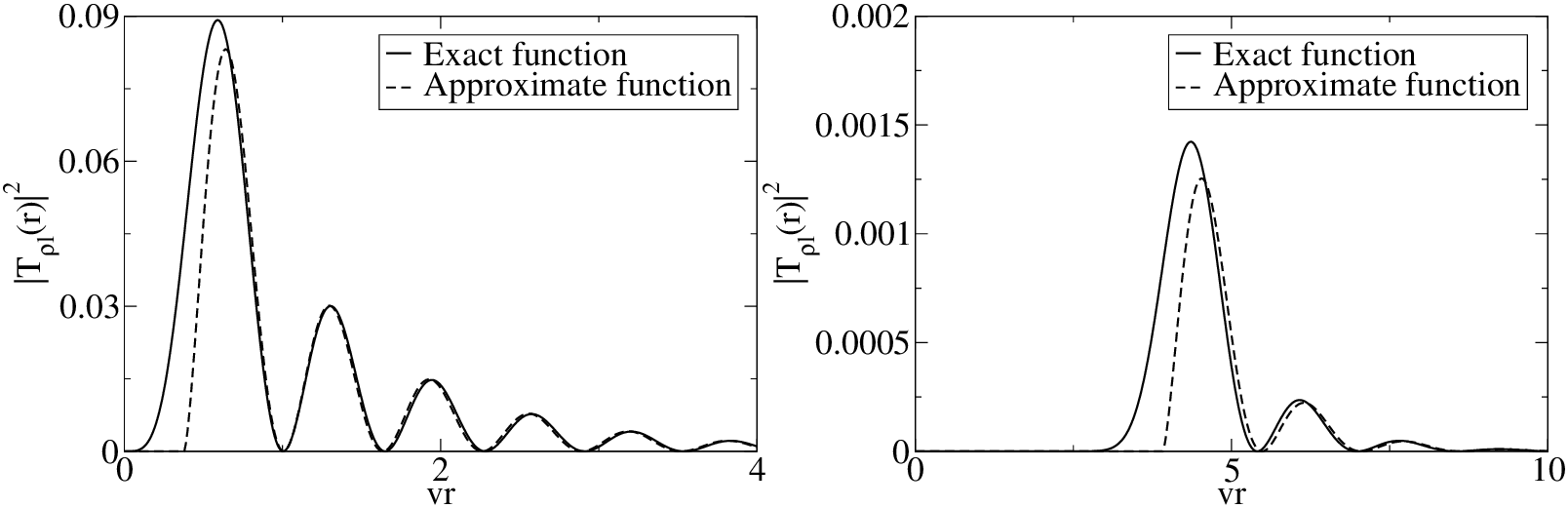}
\caption{The squared absolute values of the approximate eigenfunctions \eqref{eq:TApprox} compared to the exact eigenfunctions \eqref{eq:T} as functions of $vr$. The solid curves are the exact functions and the dashed curves are the approximate functions. In the left panel, the functions are plotted for the parameter values $\rho = 5$ and $\ell = 2$, whereas in the right panel, they are plotted for the parameter values $\rho = 2$ and $\ell = 50$.}\label{fig:Eigenfunctions}
\end{figure}

A sample of the approximate functions, as well as the corresponding exact functions, are presented in Fig.~\ref{fig:Eigenfunctions}. We have plotted the squared absolute values of these functions, since these are the quantities that enter in the physical results. For small $vr$, where the exact functions have not yet started to increase appreciably, the approximations agree with the exact functions to good accuracy. As expected, the agreement is worse in the region around $\tau = \tau_0$, where the approximate functions start to oscillate. From the first minimum in the oscillating region and on, the approximation is once again very accurate, and the agreement increases with increasing $vr$. Also, our numerical investigations indicate that the approximation becomes better as the parameters $\rho$ and $\ell$ are increased.

\subsection{Constraints on the parameter space}\label{Sec:Constraints}

In the ADD model, there is only a single free parameter, which can be taken to be the fundamental mass scale $M_*$. In our model, on the other hand, the curvature $v$ of the internal space and the position of the brane in the radial direction $\tau_b$ enter as two additional free parameters. The mass scale $M_*$ is bounded from below to $M_* \gtrsim 1 \,\, {\rm TeV}$ by the fact that no signs of quantum gravitational effects have been found in experiments up to this scale \cite{Csaki:2004ay}. On the other hand, $M_*$ cannot be much larger than $1 \,\, {\rm TeV}$ if the model is to provide a solution to the hierarchy problem. The fact that our model is an effective theory, valid only up to energy scales of the order of $M_*$, means that we should not consider values of $v$ larger than this scale.

Further constraints can be found by demanding that the model should not be in conflict with other well-established physical phenomena. In the context of the ADD model, a number of such constraints have been analyzed in Ref.~\cite{ArkaniHamed:1998nn}. In particular, strong constraints come from astrophysics and cosmology. If sufficiently light, the lightest KK mode could be produced in large numbers in high-temperature systems, such as supernovae, and carry away large amounts of energy. This could potentially alter the evolution of the system in a non-acceptable way. For the ADD model, this places important constraints on $M_*$ \cite{Cullen:1999hc}. In our model, the mass of the lightest KK mode is bounded from below by $m_1 = v/2$. If $v$ is chosen so that this mass is larger than the temperature of a supernova, which for SN1987A is about $50 \,\, {\rm MeV}$, then the constraints are completely avoided. This is achieved for $v > 100 \,\, {\rm MeV}$. Note that this bound has not been optimized, but merely gives an order-of-magnitude estimate. Of course, the bound also depends on $M_*$. However, a more detailed analysis of the exact constraints on the full parameter space is beyond the scope of this paper. As has been mentioned earlier, we only consider values of $v$ of the order of $M_*$, in order to obtain a satisfactory solution to the hierarchy problem. Since $M_*$ is of the order of $1 \,\, {\rm TeV}$, $v$ is far larger than $100 \,\, {\rm MeV}$. In the same way, we consider $\tau_b$ to be unrestricted to lie anywhere in the range $[0,\tau_{\rm max}]$, regardless of $M_*$ and $v$.

\subsection{Numerical analysis of the Kaluza--Klein spectrum}

Since it is not possible to obtain the KK spectrum of the graviton analytically, we have used a combination of numerical calculations and Weyl's asymptotic formula, as well as the form of the approximate solutions \eqref{eq:TApprox} obtained by using the WKB approximation. We assume that there are no zeros of the derivative of $T_{\rho\ell}(\tau)$ for $\tau < \tau_0$. Thus, for a given value of $\ell$, this means that any allowed value of $\rho$ has to be such as to fulfill the relation
\begin{equation}\label{eq:SpMax}
	\rho^2 \sinh^2 (\tau_{\rm max}) \geq \ell^2 - 1/4,
\end{equation}
where $\tau_{\rm max} \equiv vL$.

Performing the numerical calculations, we have found that the spectrum increases logarithmically for small $m$, {\it i.e.}, $m_n \sim \log (n)$. Comparing it with the result of Weyl's asymptotic formula, which increases as $m_n \sim \sqrt{n}$, it is in fact nearly constant at $m \approx m_1 = v/2$. This is expected, since the spectrum from Weyl's formula starts out at $m = 0$, while the true spectrum starts out at the non-zero value $m_1$. Hence, the true spectrum has to increase slower than the approximate formula in order for the two results to converge for large $n$. We have not been able to solve numerically for the spectrum up to values where the two results converge. Instead, we have resorted to solving numerically only for a manageable number, $\mathcal{O}(10^3)$, of eigenvalues and extrapolating these results up to a point where Weyl's formula is supposed to hold. This point is taken as the intersection between the extrapolated results from the numerical solution and the result from Weyl's formula. Since the spectrum is nearly constant at $m = m_1$ in the lower regime, our final result is
\begin{equation}\label{eq:ApproxSp}
	N(m^2) = \frac{\sinh^2(vL/2)}{v^2} m^2 \Theta (m-m_1),
\end{equation}
where $\Theta (x)$ is the Heaviside step function.

\section{Interactions between the graviton and the SM fields}\label{Sec:Interactions}

\subsection{The interaction Lagrangian}\label{Sec:InteractionLagrangian}

In order to study interactions between the KK modes of the graviton and the SM fields, we need the interaction terms in the action. In the general case when $h_{Mi} \neq 0$, the higher-dimensional coupling between gravity and the SM fields to first order in $h_{MN}/M_*^2$ is given by
\begin{equation}\label{eq:SInt}
	S^{\rm (int)} = \frac{1}{M_*^2} \int \ud^4 x \int \ud^2 y \, T^{MN}(x,y) h_{MN}(x,y).
\end{equation}
Because of the confinement of the SM fields to the brane at $y = y_b$, the energy-momentum tensor is $T^{MN}(x,y) = \delta_\mu^M \delta_\nu^N T^{\mu\nu}(x) \delta^{(2)}(y-y_b)$ (no summation), where $\delta^{(2)}(y)$ is the Dirac delta function and $T^{\mu\nu}(x)$ is the ordinary SM energy-momentum tensor. Inserting this expression into Eq.~\eqref{eq:SInt} and using the KK expansion \eqref{eq:KKDecomposition} yields the individual interaction terms
\begin{equation}
	S_{\rho\ell}^{\rm (int)} = \frac{1}{M_*^2} \int \ud^4x T^{\mu\nu} (x) \frac{1}{\| \psi_{\rho\ell} \|} h_{\rho\ell,\mu\nu} (x) \psi_{\rho\ell} (y_b),
\end{equation}
where we have explicitly displayed the normalization of $\psi_{\rho\ell}$. Using the relation $M_{\rm Pl}^2 = V M_*^4$, these interaction terms can be rewritten as
\begin{equation}\label{eq:SIntInd}
	S_{\rho\ell}^{\rm (int)} = \frac{c_{\rho\ell}}{{M}_{\rm Pl}} \int \ud^4x T^{\mu\nu} (x)  h_{\rho\ell,\mu\nu} (x),
\end{equation}
where
\begin{equation}
	c_{\rho\ell} \equiv \frac{\psi_{\rho\ell} (y_b) V^{1/2}}{\| \psi_{\rho\ell} \|},
\end{equation}
which are dimensionless numbers characterizing the coupling strengths. In the ADD model, the corresponding constants $c_{\bar n} = \exp \left( {\rm i} \bar n \cdot \bar y / R \right)$ are unimodular, and hence, they do not affect any physical results. In our model, the numbers $c_{\rho\ell}$ are generally not unimodular, which has the results that different KK modes have different coupling strengths and that these coupling strengths depend on the position of the brane in the internal space. However, note that the angular parts of the eigenfunctions, $\exp \left( {\rm i} \ell \varphi \right)$, are still unimodular, and thus, $c_{\rho\ell}$ only depends on the radial position of the brane and not on the angular position. From the above discussion, it follows that the constants $c_{\rho\ell}$ provide a parametrization of the difference between the ADD model and the hyperbolic disc model in the coupling strengths of the individual KK modes.

\subsection{Graviton production cross sections}

Since the interaction terms in the action \eqref{eq:SIntInd} differ from those in the ADD model only by the constant factors $c_{\rho\ell}$, the Feynman rules for the hyperbolic disc model are the same as those for the ADD model, except that the vertex factors involving KK modes of the graviton are multiplied by these same factors. The Feynman rules for the ADD model are given in Ref.~\cite{Giudice:1998ck}, and since our normalization convention agrees with theirs, the results can be used to obtain the Feynman rules for the hyperbolic disc model. Note, though, that their results are expressed in terms of the quantity $M_D$, which is related to $M_*$ as $M_D = \sqrt{2\pi} M_*$.

Since the amplitudes for processes involving KK modes of the graviton are suppressed by the Planck scale, the cross sections for production of a single KK mode are extremely low. However, at sufficiently high energies, a large number $N(m_{\rm max}^2) = N(E_{\rm cm}^2) \approx v^{-2} \sinh^2 \left( vL/2 \right) E_{\rm cm}^2$ of KK modes are kinematically available. The cross sections for production of any available KK mode are suppressed only by powers of the higher-dimensional gravitational mass scale $M_*$, which could be significantly smaller than the ordinary Planck scale. Once produced, a KK mode is extremely weakly interacting and consequently it appears as missing energy in detectors.

The cross section for production of any KK mode is obtained by summing over the kinematically available individual cross sections, {\it i.e.},
\begin{equation}
	\frac{\ud \sigma}{\ud t} = \sum_{m \leq \sqrt{s}} \frac{\ud \sigma_m}{\ud t},
\end{equation}
where $s$ and $t$ are the usual Mandelstam variables. The standard way to treat this sum is to observe that the mass splittings are small in relation to any other relevant mass scale. In the ADD model, the mass splittings are $\Delta m \sim R^{-1}$, while in our model, the mass splittings are $\Delta m \sim v \sinh^{-1} \left( vL/2 \right)$, which is small for the regions of the parameter space that we consider. Thus, the sum can be approximated to good accuracy by an integral, {\it i.e.},
\begin{equation}\label{eq:sum}
	\frac{\ud \sigma}{\ud t} \approx \int \ud m \, n(m) \frac{\ud \sigma_{m}}{\ud t},
\end{equation}
where $n(m)$ is the density of states for the KK modes. Using the approximate expression for the spectrum given in Eq.~\eqref{eq:ApproxSp}, the density of states is
\begin{equation}
	n(m) = \frac{\ud N(m^2)}{\ud m} = \delta (m-m_1) \frac{\sinh^2(vL/2)}{v^2} m^2 + \Theta (m-m_1) \frac{2\sinh^2(vL/2)}{v^2} m.
\end{equation}

\begin{figure}[htb]
\centering
\includegraphics[width=\textwidth,clip]{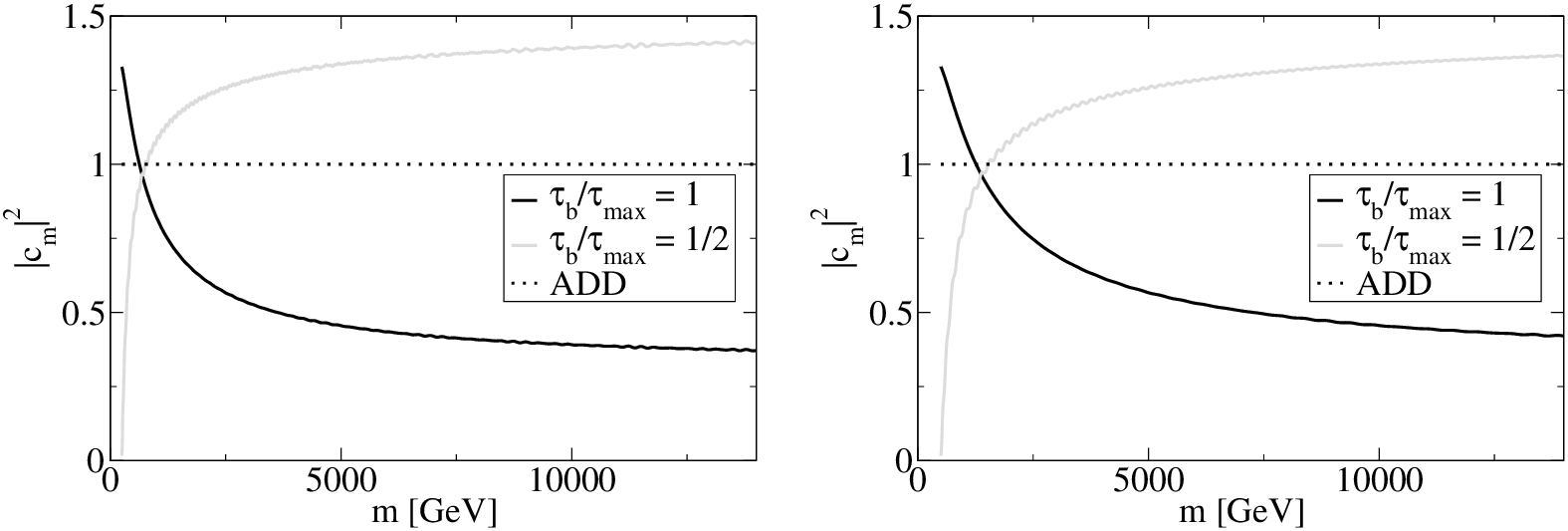}
\caption{The squared absolute values of the averaged coefficients $|c_m|^2$ as functions of $m$. The black curves correspond to the parameter value $\tau_b = \tau_{\rm max}$ and the gray ones correspond to the value $\tau_b = \tau_{\rm max}/2$. As a reference, the corresponding trivial results for the ADD model are also shown as dotted lines. In the left panel, the results are plotted for $M_* = 1 \,\, {\rm TeV}$, whereas in the right panel, they are plotted for $M_* = 2 \,\, {\rm TeV}$.}\label{fig:Couplings}
\end{figure}

There is a technical problem related to the evaluation of the integral \eqref{eq:sum}. While the differential cross section $\ud \sigma_m / \ud t$ and the density of states $n(m)$ are given as functions of $m$, the constants $c_{\rho\ell}$ are only available as functions of the KK indices $\rho$ and $\ell$, and we have no analytic relation between these quantities. Hence, in order to evaluate the integral, we need to numerically translate the constants into functions of $m$. This can be done by averaging $c_{\rho\ell}$ over the allowed values of $\ell$ for each value of $\rho$, and using the relation \eqref{eq:MRhoEll} to express $\rho$ in terms of $m$. From Eq.~\eqref{eq:SpMax}, it follows that, for fixed $\rho$, $\ell$ is restricted to the interval $[-\ell_{\rm max},\ell_{\rm max}]$, where $\ell_{\rm max} = \sqrt{\rho^2 \sinh^2 (\tau_{\rm max}) + 1/4}$. Now, in order to perform the averaging, an estimate of the density of eigenvalues for given $\rho$ and $\ell$ is needed. The approximate eigenfunctions \eqref{eq:TApprox} that we employ consist of a decaying factor and an oscillating factor $\sin[\rho (\tau - \tau_0)]$. Since the eigenvalues are determined by the positions of the zeros of $\ud T_{\rho\ell}/\ud \tau |_{\tau=\tau_b}$ as a function of $\rho$, we consider the eigenfunctions evaluated at the position of the brane, {\it i.e.}, at $\tau = \tau_b$. In addition, since $\tau_0 = {\rm arsinh} \left[ \sqrt{(\ell^2-1/4)/\rho^2} \right]$ is a slowly varying function of $\rho$ and $\ell$, we may locally consider the oscillating factor, seen as a function of $\rho$, to have a well-defined wave number equal to $\tau_b - \tau_0$. Thus, the density of zeros of $T_{\rho\ell}(\tau_b)$ as a function of $\rho$ is approximately proportional to $\tau_b - \tau_0$. We are interested in the zeros of the derivative $\ud T_{\rho\ell}/\ud \tau |_{\tau=\tau_b}$, and we assume that there is exactly one such zero between each pair of zeros of $T_{\rho\ell}(\tau_b)$. Hence, for given $\rho$ and $\ell$, the density of eigenvalues is also proportional to $\tau_b - \tau_0$. Using this result, the averaging has been performed. A sample of the resulting functions is shown in Fig.~\ref{fig:Couplings}. Note that in the resulting coefficients $c_m$, the rapid oscillations of the eigenfunctions have been washed out, leaving smooth functions.

Finally, we need to take into account that the colliding particles are protons, while the cross sections are given on the more fundamental quark level. Cross sections for such processes are calculated using the parton model. The total cross section for a high-energy hadron-hadron collision can be written as a convolution of two parton distribution functions with a hard-scattering parton-level cross section $\hat \sigma$ \cite{Sterman:2004pd},
\begin{equation}
	\sigma_{A + B \to X} (s) = \sum_{a,b} \int_0^1 \ud x_1 \int_0^1 \ud x_2 f_{a/A} (x_1,\hat{s}) f_{b/B} (x_2,\hat{s}) \hat{\sigma}_{a + b \to X}(\hat{s}),
\end{equation}
where $f_{a/A}$ is the parton distribution function for the parton $a$ in the hadron $A$, $s$ is the squared center-of-mass energy in the hadron-hadron system, $x_1$ and $x_2$ are the momentum fractions of the hadrons carried by the respective partons, and $\hat s \equiv x_1 x_2 s$ is the effective center-of-mass energy squared in the parton-parton system. In principle, the sum is to be taken over all parton species, {\it i.e.}, quarks, anti-quarks, and gluons, although the heavy quarks, \emph{i.e.}, charm, bottom, and top, are usually neglected. In this paper, we use the CTEQ6M \cite{Pumplin:2002vw} parton distribution functions.

\subsection{LHC graviton production reactions}

We consider the reactions $p+p \to {\rm jet} + G$ and $p+p \to \gamma + G$, where $G$ denotes a KK mode of the graviton. For the ADD model, the individual differential cross sections for these reactions are given in Ref.~\cite{Giudice:1998ck}. On the parton level, the reaction $p+p \to {\rm jet} + G$ consists of the three subprocesses $q + \bar q \to g + G$, $q + g \to q + G$, and $g + g \to g + G$, while the reaction $p+p \to \gamma + G$ consists of the single subprocess $q + \bar q \to \gamma + G$. Since each of these subprocesses includes a single vertex involving a graviton, the cross sections for our model are obtained by multiplying the results for the ADD model by the constant factors $|c_{\rho\ell}|^2$.

As mentioned above, the produced gravitons are very weakly interacting, and hence, they appear as missing energy in detectors. Thus, the observed reactions are $p+p \to {\rm jet} + \slc{E}$ and $p+p \to \gamma + \slc{E}$, respectively, where $\slc{E}$ denotes the missing energy.

\section{Numerical analysis of signals}\label{Sec:NumericalAnalysisOfSignals}

In this section, we present the predictions of our model. For both cases of jet and photon production, we have calculated differential cross sections with respect to $\cos(\theta)$, where $\theta$ is the angle between the proton beam and the outgoing jet/photon, as well as with respect to $p_{\rm T, jet/\gamma}$, which is the momentum of the jet or photon perpendicular to the beam. All of our results are presented for $\sqrt{s} = 14 \,\, {\rm TeV}$.

Since we consider high-energy processes, the partons are approximated as being massless. As there are massless particles in the final states in both of the processes that we study, the expressions for the cross sections suffer from collinear divergences in the limit of zero transverse momentum of these particles. In order to avoid these singularities, we impose a lower cut-off on the transverse momentum, $p_{\rm T,\gamma/{\rm jet}} \geq p_{\rm T}^{\rm min}$. This cut-off also serves to increase the signal-to-background ratio. Because of the finite size of the detector, there is an upper cut-off on the longitudinal rapidity (or pseudorapidity) $\eta = {\rm artanh} \left[ \cos (\theta) \right]$ of these outgoing particles, {\it i.e.}, $|n_{\gamma/{\rm jet}}| \leq \eta_{\rm max}$. We have used the value $\eta_{\rm max} = 2.5$ for all measurements considered \cite{ATLAS:1999fr}.

It is important to take into consideration the fact that the theory is an effective one only, which is supposed to break down at large energies, above some cut-off scale. Without knowledge of the UV completion of the effective theory, it is not possible to determine this cut-off scale exactly. We follow Ref.~\cite{Giudice:1998ck} and trust our results only up to the mass scale $M_D = \sqrt{2 \pi} M_*$. We also follow their method of analyzing the validity of the results, {\it i.e.}, by computing cross sections that are set to zero for ${\hat s} > M_D^2$, and compare these to the naive results. In regions where the results agree, almost all of the contributions to the cross sections come from subprocesses with an effective center-of-mass energy lower than the fundamental mass scale, and hence, these results can be trusted. These regions depend on the chosen set of parameters, and differ between jet and photon reactions. In general, the results become better for higher $M_*$, but at the same time the cross sections decrease. Hence, we need to make a trade-off between these two competing effects.

As discussed in Sec.~\ref{Sec:TheHyperbolicDiscModel}, we only consider values of $v$ of the order of $M_*$. However, for $v = M_*$, it is difficult to to obtain valid results for the effective model. Hence, for all the results presented, we have have set $v = M_*/2$, in which case it is possible to obtain valid results. Complementary to these considerations of internal spaces with large curvature, internal spaces with small curvature have been considered in Ref.~\cite{Giudice:2004mg}.

The position of the brane in the radial direction is not constrained. However, as discussed in Sec.~\ref{Sec:TheHyperbolicDiscModel}, in the case that the brane is positioned at the center of the disc, only the $\ell = 0$ KK modes couple to the SM fields. This means that, while the number of kinematically available KK modes is typically of the order of $10^{17}$ for the cases that we consider, in this special case this number is effectively reduced to a number of the order of $100$. Hence, only a negligibly small fraction of the KK modes are effectively available in this case, and the signal will be far too weak to be observable at the LHC. In order to give a better sense for the range of rates possible in the model, it could still be interesting to obtain quantitative results for this special case. However, this analysis is complicated by the fact that the effective mass splittings between the active KK modes become too large to allow us to employ the approximation~(\ref{eq:sum}) when calculating the total cross section, and thus, we have not performed any such calculations. For each fixed set of values for the rest of the parameters in the model, we have presented our results for two different values of the position of the brane, at $\tau_b = \tau_{\rm max}$ and at $\tau_b = \tau_{\rm max}/2$.

Note also that, as our calculations are performed to leading order only, we do not take final state radiation into account.

\subsection{$p + p \to {\rm jet} + \slc{E}$}

For the jet, we have chosen a transverse momentum cut-off $p_{\rm T}^{\rm min} = 750 \,\, {\rm GeV}$. The results are presented for two different values of the fundamental mass scale, $M_* = 1.5 \,\, {\rm TeV}$ and $M_* = 2 \,\, {\rm TeV}$. The main background comes from the processes $p + p \to {\rm jet} + Z$ and $p + p \to {\rm jet} + W$, with the $Z$ decaying into a neutrino-antineutrino pair and the $W$ decaying into a neutrino and a lepton, respectively \cite{Vacavant:2001sd}. In the case of $W$ production, the background can be distinguished from the signal if a lepton in the outgoing state is observed. Taking this into account, simulations of the background using PYTHIA \cite{Sjostrand:2006} shows that the background from $W$ production is small in comparison to the background from $Z$ production, and hence, we have only performed accurate simulations for $Z$ production. 

\begin{figure}[htb]
\centering
\includegraphics[width=\textwidth,clip]{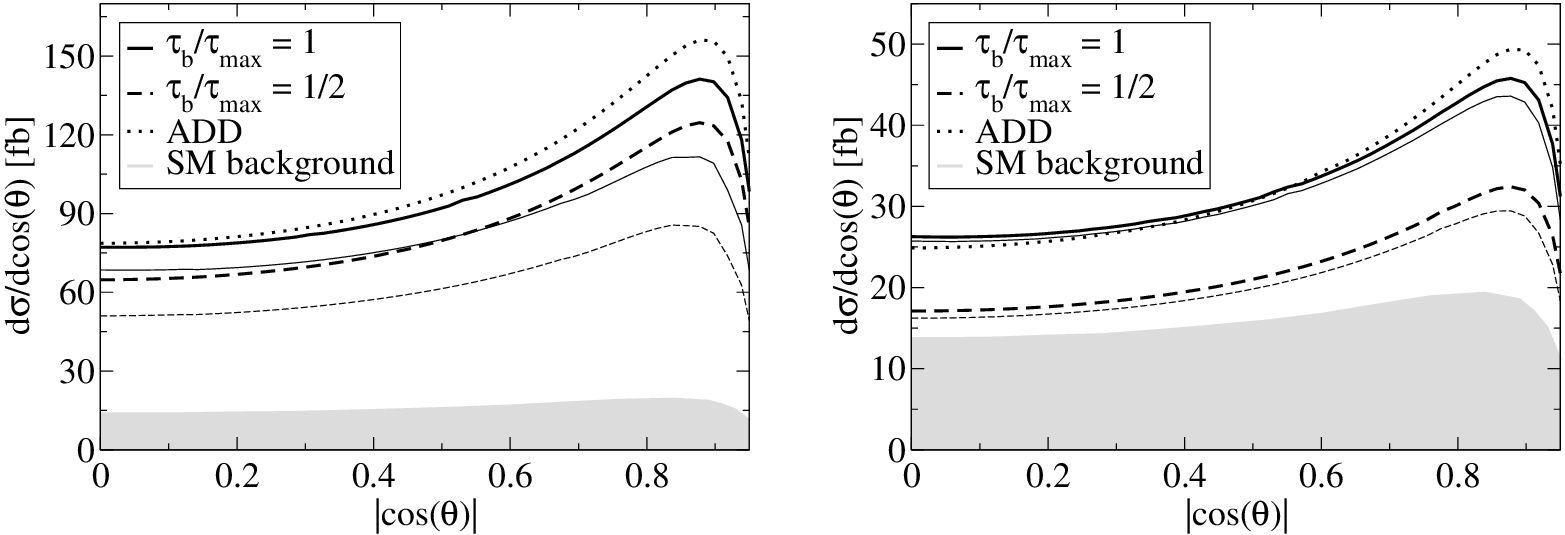}
\caption{The differential cross section for graviton plus jet production with respect to $\cos (\theta)$. The solid curves correspond to the parameter value $\tau_b = \tau_{\rm max}$ and the dashed curves correspond to the value $\tau_b = \tau_{\rm max}/2$. For each of these two values, the thick lines are the naive cross sections and the corresponding thin lines are the truncated ones. The dotted lines are the corresponding results for the ADD model with the same value for $M_*$, and the gray shaded area is the SM background. In the left panel, the results are plotted for $M_* = 1.5 \,\, {\rm TeV}$, whereas in the right panel, they are plotted for $M_* = 2 \,\, {\rm TeV}$.}\label{fig:JetCos}
\end{figure}

\begin{figure}[htb]
\centering
\includegraphics[width=\textwidth,clip]{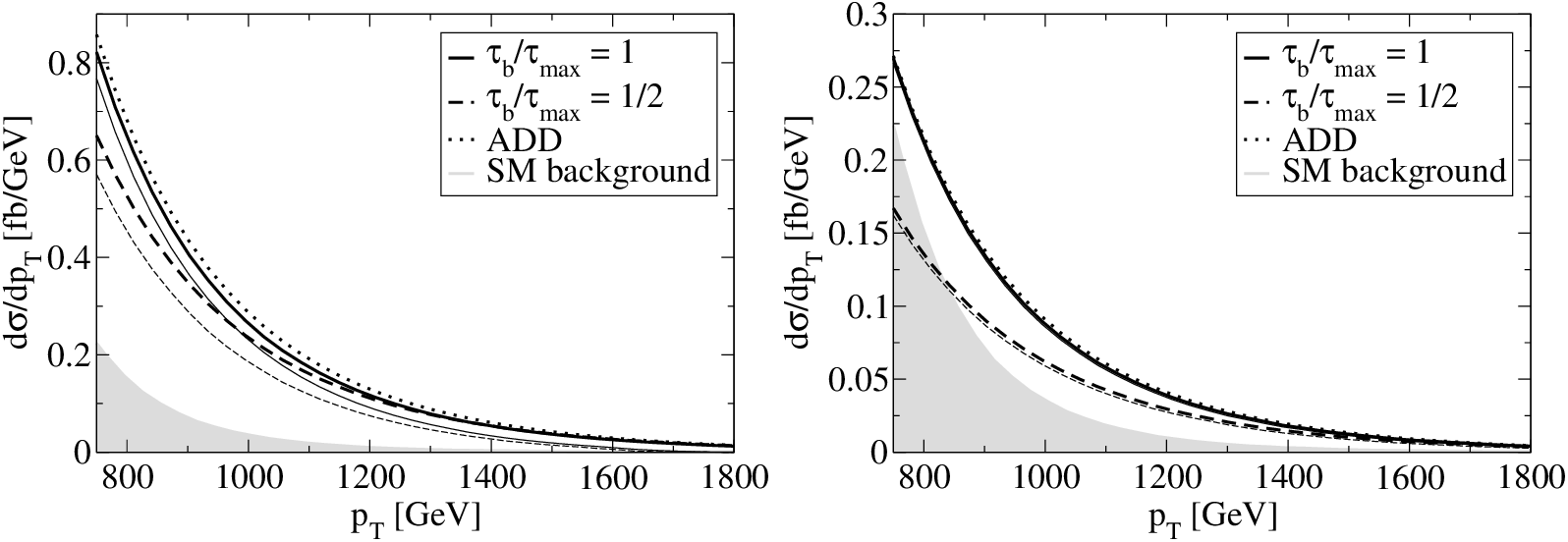}
\caption{The differential cross section for graviton plus jet production with respect to $p_{\rm T}$. The solid curves correspond to the parameter value $\tau_b = \tau_{\rm max}$ and the dashed curves correspond to the value $\tau_b = \tau_{\rm max}/2$. For each of these two values, the thick lines are the naive cross sections and the corresponding thin lines are the truncated ones. In the left panel, the results are plotted for $M_* = 1.5 \,\, {\rm TeV}$ and in the right panel, they are plotted for $M_* = 2 \,\, {\rm TeV}$.}\label{fig:JetPT}
\end{figure}

The differential cross sections $\ud \sigma / \ud \! \cos(\theta)$ and $\ud \sigma / \ud p_{\rm T,jet}$ are given in Figs.~\ref{fig:JetCos} and \ref{fig:JetPT}, respectively. As a reference, we have also plotted the corresponding cross sections for the ADD model with the same value for $M_*$. The cross sections for our model resemble those of the ADD model and are of the same order of magnitude. For $\tau_b = \tau_{\rm max}$, the result is almost indistinguishable from the ADD results for both $M_* = 1.5 \,\, {\rm TeV}$ and $M_* = 2 \,\, {\rm TeV}$. As expected, the cross sections decrease with increasing $M_*$, while the discrepancy between the naive and the truncated cross sections increase with decreasing $M_*$. Also, the signals have the same behavior as the background. For all the demonstrated results, the signals are larger than the background, although not by much for $M_* = 2 \,\, {\rm TeV}$. However, for $M_* = 1.5 \,\, {\rm TeV}$, the discrepancies between the naive and the truncated cross sections are quite large.

Note that there is also a difference between results for different values of $\tau_b$. For both values of $M_*$ shown, the cross sections are larger for $\tau_b = \tau_{\rm max}$.

For $M_* = 1.5 \,\, {\rm TeV}$, the integrated cross sections are of the order of $200 \,\, {\rm fb}$. Thus, for an integrated luminosity at the LHC of $10 \,\, {\rm fb}^{-1}$ or $100 \,\, {\rm fb}^{-1}$, the expected number of events is of the order of $2000$ or $20000$, respectively, whereas for $M_* = 2 \,\, {\rm TeV}$, the cross sections are of the order of $50 \,\, {\rm fb}$, and the corresponding number of events are of the order of $500$ or $5000$.

\subsection{$p + p \to \gamma + \slc{E}$}

For the photon, we have chosen a transverse momentum cut-off $p_{\rm T}^{\rm min} = 300 \,\, {\rm GeV}$. The results are presented for $M_* = 1 \,\, {\rm TeV}$ and $M_* = 1.5 \,\, {\rm TeV}$. The main background is analogous to the background for the jet production process, {\it i.e.}, coming from the processes $p + p \to \gamma + Z$ and $p + p \to \gamma + W$ \cite{Vacavant:2001sd}. In the same way as in the jet case, the background from $Z$ production is dominant, and we have only considered this contribution to the background.

\begin{figure}[htb]
\centering
\includegraphics[width=\textwidth,clip]{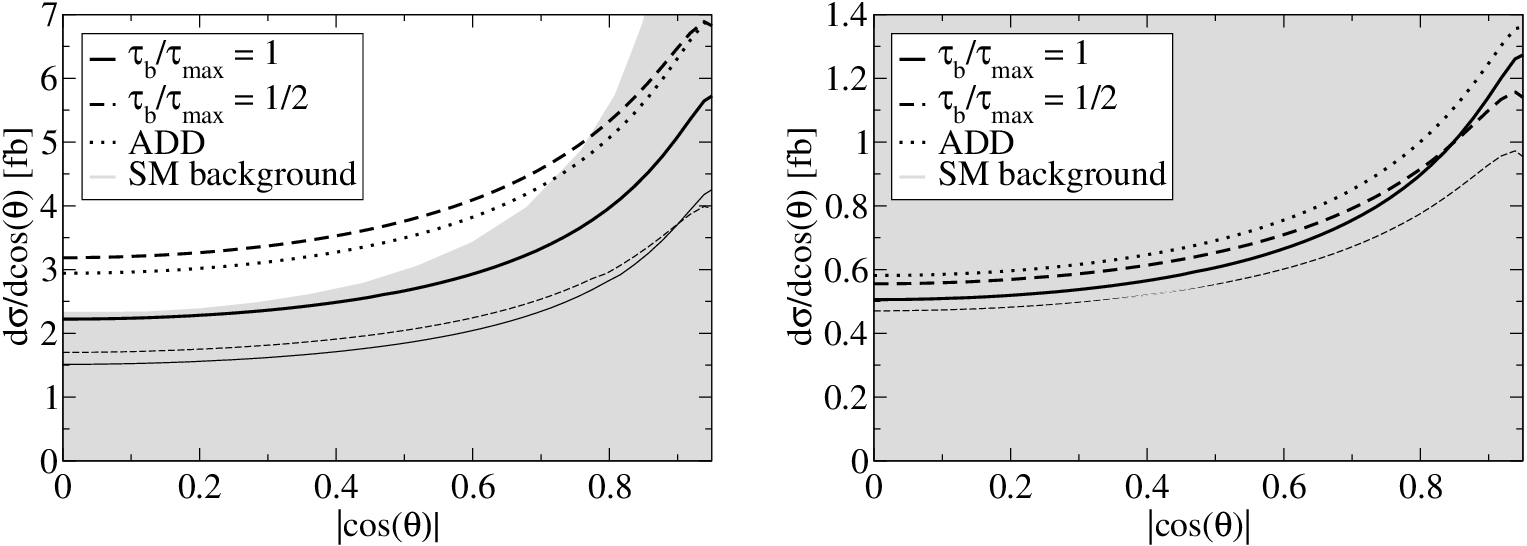}
\caption{The differential cross section for graviton plus photon production with respect to $\cos (\theta)$. The solid curves correspond to the parameter value $\tau_b = \tau_{\rm max}$ and the dashed curves correspond to the value $\tau_b = \tau_{\rm max}/2$. For each of these two values, the thick lines are the naive cross sections and the corresponding thin lines are the truncated ones. In the left panel, the results are plotted for $M_* = 1 \,\, {\rm TeV}$, whereas in the right panel, they are plotted for $M_* = 1.5 \,\, {\rm TeV}$.}\label{fig:PhotonCos}
\end{figure}

\begin{figure}[htb]
\centering
\includegraphics[width=\textwidth,clip]{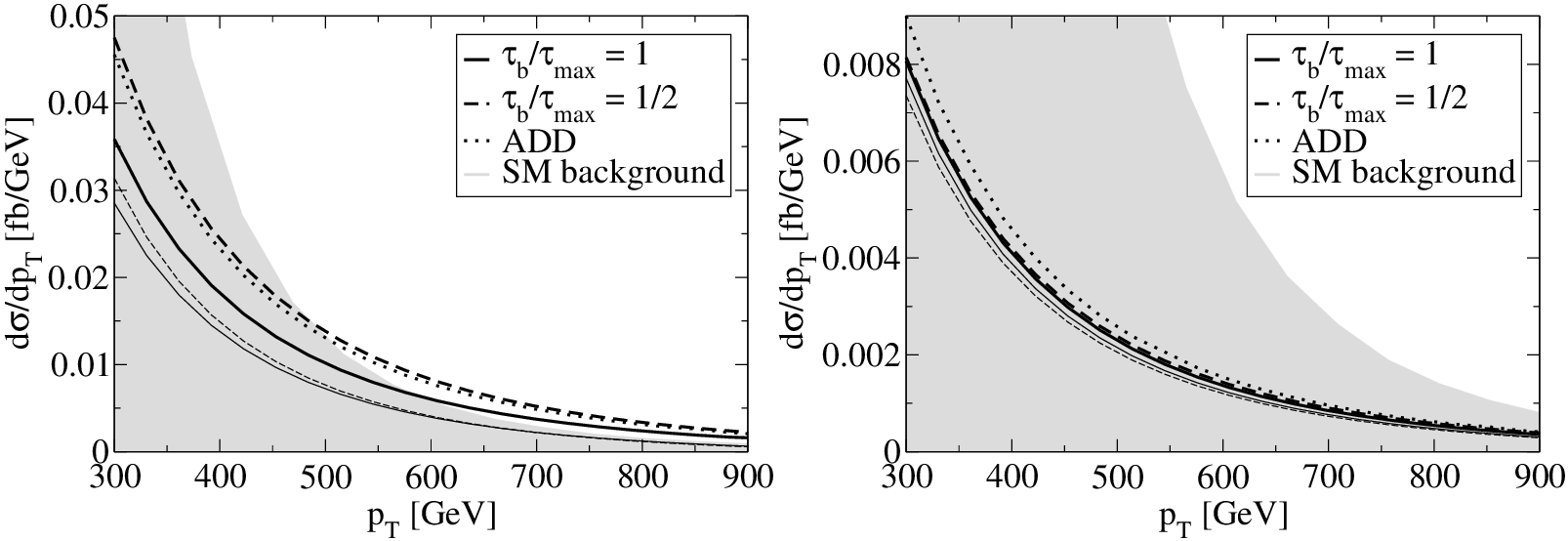}
\caption{The differential cross section for graviton plus photon production with respect to $p_{\rm T}$. The solid curves correspond to the parameter value $\tau_b = \tau_{\rm max}$ and the dashed curves correspond to the value $\tau_b = \tau_{\rm max}/2$. For each of these two values, the thick lines are the naive cross sections and the corresponding thin lines are the truncated ones. In the left panel, the results are plotted for $M_* = 1 \,\, {\rm TeV}$, whereas in the right panel, they are plotted for $M_* = 1.5 \,\, {\rm TeV}$.}\label{fig:PhotonPT}
\end{figure}

The differential cross sections $\ud \sigma / \ud \! \cos(\theta)$ and $\ud \sigma / \ud p_{{\rm T},\gamma}$ are presented in Figs.~\ref{fig:PhotonCos} and \ref{fig:PhotonPT}, respectively. As for the case of jet production, the cross sections resemble those of the ADD model and also have a similar behavior to the background. In this case, the results for $\tau_b = \tau_{\rm max}/2$ are very similar to the results for the ADD model. In comparison to the jet production case, it is much more difficult to find a region where the effective theory is valid and the signal is not much smaller than the background. In particular, in the case that $M_* = 1.5 \,\, {\rm TeV}$, the background is much larger than the signal.

For $M_* = 1 \,\, {\rm TeV}$, the integrated cross sections are of the order of $10 \,\, {\rm fb}$. Thus for an integrated luminosity at the LHC of $10 \,\, {\rm fb}^{-1}$ or $100 \,\, {\rm fb}^{-1}$, the expected number of events is of the order of $100$ or $1000$, respectively, whereas for $M_* = 1.5 \,\, {\rm TeV}$, the cross sections are of the order of $1 \,\, {\rm fb}$, and the corresponding number of events are of the order of $10$ or $100$. Comparing to the jet production case, the cross sections are much smaller, about two orders of magnitude for $M_* = 1.5 \,\, {\rm TeV}$.

\section{Summary and conclusions}\label{Sec:SummaryAndConclusions}

In this paper, we have considered a model for large extra dimensions based on hyperbolic geometry. More specifically, the internal space has the geometry of a hyperbolic disc with constant curvature. This model is in some sense a generalization of the ADD model. Its main advantages are that it provides a more satisfactory solution to the hierarchy problem than the ADD model, and that constraints from astrophysics on the size of the extra dimensions are avoided. 

We have investigated the possible experimental signatures of the model at the LHC. The two main reactions that could be relevant for the LHC are $p + p \to {\rm jet} + G$ and $p + p \to \gamma + G$, where $G$ denotes a KK mode of the graviton. As the KK modes are extremely weakly interacting, the corresponding amplitudes being suppressed by the Planck scale, the gravitons produced in these reactions are not detected, but appear as missing energy in detectors. Since there is no known analytic form for the KK spectrum when the internal space is hyperbolic, we have employed a combination of approximations and numerical investigations to obtain our results. 

We have found that some regions of the parameter space could be probed by the LHC, using the jet production channel. In this reaction, the integrated cross sections are typically of the order of $100 \,\, {\rm fb}$ for the cases that have been studied. For the reaction involving the production of a photon, the discovery potential is significantly weaker than in the former case. For large $M_*$, the background is much larger than the signal, while for smaller $M_*$, the applicability of the effective theory breaks down and the predictions cannot be trusted. Also, the integrated cross sections are about two orders of magnitude smaller than for the jet production case, typically of the order of $1 \,\, {\rm fb}$. For both cases of jet and photon production, the cross sections have the same qualitative behavior and are of the same order of magnitude as the cross sections for the ADD model with the same value for $M_*$. In fact, the signals of our model are in some cases indistinguishable from signals of the ADD model. In addition, the parameter space of our model is much larger than that of the ADD model, our model having three free parameters with only weak experimental constraints.

A novel feature of our model, in comparison to the ADD model, is that its physical predictions depend on the position of the brane in the internal space. In particular, in the case that the brane is placed at the center of the disc, {\it i.e.}, at $\tau = 0$, most of the coupling constants between the KK modes of the graviton and SM fields vanish. This has the result that in this case the experimental signatures of the model would be far too weak to be observable in collider experiments.

In conclusion, we have found that the most promising channel for the detection of hyperbolic extra dimensions at the LHC is the production of a KK mode of the graviton together with a single hadronic jet. In the case when the fundamental mass scale $M_*$ as well as the curvature $v$ are of the order of $1 \,\, {\rm TeV}$, a solution to the hierarchy problem is obtained. Depending on the position of the brane in the radial direction $\tau_b$, it may also be possible to obtain an observable signal at the LHC.

\begin{acknowledgments}

We would like to thank Tomas H{\"a}llgren, David Milstead, and Philippe Mermod for useful
discussions.

This work was supported by the Swedish Research Council (Vetenskapsr{\aa}det),
contract no. 621-2005-3588 [T.O.] and the Royal Swedish Academy of
Sciences (KVA) [T.O.].

\end{acknowledgments}


\begin{thebibliography}{35}
\expandafter\ifx\csname natexlab\endcsname\relax\def\natexlab#1{#1}\fi
\expandafter\ifx\csname bibnamefont\endcsname\relax
  \def\bibnamefont#1{#1}\fi
\expandafter\ifx\csname bibfnamefont\endcsname\relax
  \def\bibfnamefont#1{#1}\fi
\expandafter\ifx\csname citenamefont\endcsname\relax
  \def\citenamefont#1{#1}\fi
\expandafter\ifx\csname url\endcsname\relax
  \def\url#1{\texttt{#1}}\fi
\expandafter\ifx\csname urlprefix\endcsname\relax\def\urlprefix{URL }\fi
\providecommand{\bibinfo}[2]{#2}
\providecommand{\eprint}[2][]{\url{#2}}

\bibitem[{\citenamefont{Kaluza}(1921)}]{Kaluza:1921tu}
\bibinfo{author}{\bibfnamefont{T.}~\bibnamefont{Kaluza}},
  \bibinfo{journal}{Sitzungsber. Preuss. Akad. Wiss. Berlin (Math. Phys. )}
  \textbf{\bibinfo{volume}{1921}}, \bibinfo{pages}{966} (\bibinfo{year}{1921}).

\bibitem[{\citenamefont{Klein}(1926)}]{Klein:1926tv}
\bibinfo{author}{\bibfnamefont{O.}~\bibnamefont{Klein}}, \bibinfo{journal}{Z.
  Phys.} \textbf{\bibinfo{volume}{37}}, \bibinfo{pages}{895}
  (\bibinfo{year}{1926}).

\bibitem[{\citenamefont{Arkani-Hamed et~al.}(1998)\citenamefont{Arkani-Hamed,
  Dimopoulos, and Dvali}}]{ArkaniHamed:1998rs}
\bibinfo{author}{\bibfnamefont{N.}~\bibnamefont{Arkani-Hamed}},
  \bibinfo{author}{\bibfnamefont{S.}~\bibnamefont{Dimopoulos}},
  \bibnamefont{and} \bibinfo{author}{\bibfnamefont{G.~R.} \bibnamefont{Dvali}},
  \bibinfo{journal}{Phys. Lett.} \textbf{\bibinfo{volume}{B429}},
  \bibinfo{pages}{263} (\bibinfo{year}{1998}), \eprint{hep-ph/9803315}.

\bibitem[{\citenamefont{Antoniadis et~al.}(1998)\citenamefont{Antoniadis,
  Arkani-Hamed, Dimopoulos, and Dvali}}]{Antoniadis:1998ig}
\bibinfo{author}{\bibfnamefont{I.}~\bibnamefont{Antoniadis}},
  \bibinfo{author}{\bibfnamefont{N.}~\bibnamefont{Arkani-Hamed}},
  \bibinfo{author}{\bibfnamefont{S.}~\bibnamefont{Dimopoulos}},
  \bibnamefont{and} \bibinfo{author}{\bibfnamefont{G.~R.} \bibnamefont{Dvali}},
  \bibinfo{journal}{Phys. Lett.} \textbf{\bibinfo{volume}{B436}},
  \bibinfo{pages}{257} (\bibinfo{year}{1998}), \eprint{hep-ph/9804398}.

\bibitem[{\citenamefont{Kaloper et~al.}(2000)\citenamefont{Kaloper,
  March-Russell, Starkman, and Trodden}}]{Kaloper:2000jb}
\bibinfo{author}{\bibfnamefont{N.}~\bibnamefont{Kaloper}},
  \bibinfo{author}{\bibfnamefont{J.}~\bibnamefont{March-Russell}},
  \bibinfo{author}{\bibfnamefont{G.~D.} \bibnamefont{Starkman}},
  \bibnamefont{and} \bibinfo{author}{\bibfnamefont{M.}~\bibnamefont{Trodden}},
  \bibinfo{journal}{Phys. Rev. Lett.} \textbf{\bibinfo{volume}{85}},
  \bibinfo{pages}{928} (\bibinfo{year}{2000}), \eprint{hep-ph/0002001}.

\bibitem[{\citenamefont{Randall and
  Sundrum}(1999{\natexlab{a}})}]{Randall:1999vf}
\bibinfo{author}{\bibfnamefont{L.}~\bibnamefont{Randall}} \bibnamefont{and}
  \bibinfo{author}{\bibfnamefont{R.}~\bibnamefont{Sundrum}},
  \bibinfo{journal}{Phys. Rev. Lett.} \textbf{\bibinfo{volume}{83}},
  \bibinfo{pages}{4690} (\bibinfo{year}{1999}{\natexlab{a}}),
  \eprint{hep-th/9906064}.

\bibitem[{\citenamefont{Randall and
  Sundrum}(1999{\natexlab{b}})}]{Randall:1999ee}
\bibinfo{author}{\bibfnamefont{L.}~\bibnamefont{Randall}} \bibnamefont{and}
  \bibinfo{author}{\bibfnamefont{R.}~\bibnamefont{Sundrum}},
  \bibinfo{journal}{Phys. Rev. Lett.} \textbf{\bibinfo{volume}{83}},
  \bibinfo{pages}{3370} (\bibinfo{year}{1999}{\natexlab{b}}),
  \eprint{hep-ph/9905221}.

\bibitem[{\citenamefont{Giudice et~al.}(1999)\citenamefont{Giudice, Rattazzi,
  and Wells}}]{Giudice:1998ck}
\bibinfo{author}{\bibfnamefont{G.~F.} \bibnamefont{Giudice}},
  \bibinfo{author}{\bibfnamefont{R.}~\bibnamefont{Rattazzi}}, \bibnamefont{and}
  \bibinfo{author}{\bibfnamefont{J.~D.} \bibnamefont{Wells}},
  \bibinfo{journal}{Nucl. Phys.} \textbf{\bibinfo{volume}{B544}},
  \bibinfo{pages}{3} (\bibinfo{year}{1999}), \eprint{hep-ph/9811291}.

\bibitem[{\citenamefont{Han et~al.}(1999)\citenamefont{Han, Lykken, and
  Zhang}}]{Han:1998sg}
\bibinfo{author}{\bibfnamefont{T.}~\bibnamefont{Han}},
  \bibinfo{author}{\bibfnamefont{J.~D.} \bibnamefont{Lykken}},
  \bibnamefont{and} \bibinfo{author}{\bibfnamefont{R.-J.} \bibnamefont{Zhang}},
  \bibinfo{journal}{Phys. Rev.} \textbf{\bibinfo{volume}{D59}},
  \bibinfo{pages}{105006} (\bibinfo{year}{1999}), \eprint{hep-ph/9811350}.

\bibitem[{\citenamefont{Mirabelli et~al.}(1999)\citenamefont{Mirabelli,
  Perelstein, and Peskin}}]{Mirabelli:1998rt}
\bibinfo{author}{\bibfnamefont{E.~A.} \bibnamefont{Mirabelli}},
  \bibinfo{author}{\bibfnamefont{M.}~\bibnamefont{Perelstein}},
  \bibnamefont{and} \bibinfo{author}{\bibfnamefont{M.~E.}
  \bibnamefont{Peskin}}, \bibinfo{journal}{Phys. Rev. Lett.}
  \textbf{\bibinfo{volume}{82}}, \bibinfo{pages}{2236} (\bibinfo{year}{1999}),
  \eprint{hep-ph/9811337}.

\bibitem[{\citenamefont{Hewett}(1999)}]{Hewett:1998sn}
\bibinfo{author}{\bibfnamefont{J.~L.} \bibnamefont{Hewett}},
  \bibinfo{journal}{Phys. Rev. Lett.} \textbf{\bibinfo{volume}{82}},
  \bibinfo{pages}{4765} (\bibinfo{year}{1999}), \eprint{hep-ph/9811356}.

\bibitem[{\citenamefont{Nussinov and Shrock}(1999)}]{Nussinov:1998jt}
\bibinfo{author}{\bibfnamefont{S.}~\bibnamefont{Nussinov}} \bibnamefont{and}
  \bibinfo{author}{\bibfnamefont{R.}~\bibnamefont{Shrock}},
  \bibinfo{journal}{Phys. Rev.} \textbf{\bibinfo{volume}{D59}},
  \bibinfo{pages}{105002} (\bibinfo{year}{1999}), \eprint{hep-ph/9811323}.

\bibitem[{\citenamefont{Rizzo}(1999)}]{Rizzo:1998fm}
\bibinfo{author}{\bibfnamefont{T.~G.} \bibnamefont{Rizzo}},
  \bibinfo{journal}{Phys. Rev.} \textbf{\bibinfo{volume}{D59}},
  \bibinfo{pages}{115010} (\bibinfo{year}{1999}), \eprint{hep-ph/9901209}.

\bibitem[{\citenamefont{Cheung and Keung}(1999)}]{Cheung:1999zw}
\bibinfo{author}{\bibfnamefont{K.-M.} \bibnamefont{Cheung}} \bibnamefont{and}
  \bibinfo{author}{\bibfnamefont{W.-Y.} \bibnamefont{Keung}},
  \bibinfo{journal}{Phys. Rev.} \textbf{\bibinfo{volume}{D60}},
  \bibinfo{pages}{112003} (\bibinfo{year}{1999}), \eprint{hep-ph/9903294}.

\bibitem[{\citenamefont{Balazs et~al.}(1999)\citenamefont{Balazs, He, Repko,
  Yuan, and Dicus}}]{Balazs:1999ge}
\bibinfo{author}{\bibfnamefont{C.}~\bibnamefont{Balazs}},
  \bibinfo{author}{\bibfnamefont{H.-J.} \bibnamefont{He}},
  \bibinfo{author}{\bibfnamefont{W.~W.} \bibnamefont{Repko}},
  \bibinfo{author}{\bibfnamefont{C.~P.} \bibnamefont{Yuan}}, \bibnamefont{and}
  \bibinfo{author}{\bibfnamefont{D.~A.} \bibnamefont{Dicus}},
  \bibinfo{journal}{Phys. Rev. Lett.} \textbf{\bibinfo{volume}{83}},
  \bibinfo{pages}{2112} (\bibinfo{year}{1999}), \eprint{hep-ph/9904220}.

\bibitem[{\citenamefont{Leblond}(2001)}]{Leblond:2001ex}
\bibinfo{author}{\bibfnamefont{F.}~\bibnamefont{Leblond}},
  \bibinfo{journal}{Phys. Rev.} \textbf{\bibinfo{volume}{D64}},
  \bibinfo{pages}{045016} (\bibinfo{year}{2001}), \eprint{hep-ph/0104273}.

\bibitem[{\citenamefont{Bauer et~al.}(2007)\citenamefont{Bauer, H{\"a}llgren,
  and Seidl}}]{Bauer:2006pf}
\bibinfo{author}{\bibfnamefont{F.}~\bibnamefont{Bauer}},
  \bibinfo{author}{\bibfnamefont{T.}~\bibnamefont{H{\"a}llgren}},
  \bibnamefont{and} \bibinfo{author}{\bibfnamefont{G.}~\bibnamefont{Seidl}},
  \bibinfo{journal}{Nucl. Phys.} \textbf{\bibinfo{volume}{B781}},
  \bibinfo{pages}{32} (\bibinfo{year}{2007}), \eprint{hep-th/0608176}.

\bibitem[{\citenamefont{Starkman
  et~al.}(2001{\natexlab{a}})\citenamefont{Starkman, Stojkovic, and
  Trodden}}]{Starkman:2000dy}
\bibinfo{author}{\bibfnamefont{G.~D.} \bibnamefont{Starkman}},
  \bibinfo{author}{\bibfnamefont{D.}~\bibnamefont{Stojkovic}},
  \bibnamefont{and} \bibinfo{author}{\bibfnamefont{M.}~\bibnamefont{Trodden}},
  \bibinfo{journal}{Phys. Rev.} \textbf{\bibinfo{volume}{D63}},
  \bibinfo{pages}{103511} (\bibinfo{year}{2001}{\natexlab{a}}),
  \eprint{hep-th/0012226}.

\bibitem[{\citenamefont{Starkman
  et~al.}(2001{\natexlab{b}})\citenamefont{Starkman, Stojkovic, and
  Trodden}}]{Starkman:2001xu}
\bibinfo{author}{\bibfnamefont{G.~D.} \bibnamefont{Starkman}},
  \bibinfo{author}{\bibfnamefont{D.}~\bibnamefont{Stojkovic}},
  \bibnamefont{and} \bibinfo{author}{\bibfnamefont{M.}~\bibnamefont{Trodden}},
  \bibinfo{journal}{Phys. Rev. Lett.} \textbf{\bibinfo{volume}{87}},
  \bibinfo{pages}{231303} (\bibinfo{year}{2001}{\natexlab{b}}),
  \eprint{hep-th/0106143}.

\bibitem[{\citenamefont{Bauer}(2006)}]{Bauer:2006ti}
\bibinfo{author}{\bibfnamefont{F.}~\bibnamefont{Bauer}} (\bibinfo{year}{2006}),
  \eprint{hep-th/0610178}.

\bibitem[{\citenamefont{Arkani-Hamed et~al.}(1999)\citenamefont{Arkani-Hamed,
  Dimopoulos, and Dvali}}]{ArkaniHamed:1998nn}
\bibinfo{author}{\bibfnamefont{N.}~\bibnamefont{Arkani-Hamed}},
  \bibinfo{author}{\bibfnamefont{S.}~\bibnamefont{Dimopoulos}},
  \bibnamefont{and} \bibinfo{author}{\bibfnamefont{G.~R.} \bibnamefont{Dvali}},
  \bibinfo{journal}{Phys. Rev.} \textbf{\bibinfo{volume}{D59}},
  \bibinfo{pages}{086004} (\bibinfo{year}{1999}), \eprint{hep-ph/9807344}.

\bibitem[{\citenamefont{Arkani-Hamed and Schwartz}(2004)}]{ArkaniHamed:2003vb}
\bibinfo{author}{\bibfnamefont{N.}~\bibnamefont{Arkani-Hamed}}
  \bibnamefont{and} \bibinfo{author}{\bibfnamefont{M.~D.}
  \bibnamefont{Schwartz}}, \bibinfo{journal}{Phys. Rev.}
  \textbf{\bibinfo{volume}{D69}}, \bibinfo{pages}{104001}
  (\bibinfo{year}{2004}), \eprint{hep-th/0302110}.

\bibitem[{\citenamefont{Chavel}(1984)}]{Chavel:1984}
\bibinfo{author}{\bibfnamefont{I.}~\bibnamefont{Chavel}},
  \emph{\bibinfo{title}{Eigenvalues in Riemannian Geometry}}
  (\bibinfo{publisher}{Academic Press}, \bibinfo{year}{1984}).

\bibitem[{\citenamefont{Weyl}(1912)}]{Weyl:1912}
\bibinfo{author}{\bibfnamefont{H.}~\bibnamefont{Weyl}}, \bibinfo{journal}{Math.
  Ann.} \textbf{\bibinfo{volume}{71}}, \bibinfo{pages}{441}
  (\bibinfo{year}{1912}).

\bibitem[{\citenamefont{Erd{\'e}lyi et~al.}(1953)\citenamefont{Erd{\'e}lyi,
  Magnus, Oberhettinger, and Tricomi}}]{Erdelyi:1953}
\bibinfo{author}{\bibfnamefont{A.}~\bibnamefont{Erd{\'e}lyi}},
  \bibinfo{author}{\bibfnamefont{W.}~\bibnamefont{Magnus}},
  \bibinfo{author}{\bibfnamefont{F.}~\bibnamefont{Oberhettinger}},
  \bibnamefont{and} \bibinfo{author}{\bibfnamefont{G.}~\bibnamefont{Tricomi}},
  \emph{\bibinfo{title}{Higher Transcendental Functions{\rm,~Vol.~1,}}}
  (\bibinfo{publisher}{McGraw-Hill Book Company}, \bibinfo{year}{1953}).

\bibitem[{\citenamefont{Zaldarriaga et~al.}(1997)\citenamefont{Zaldarriaga,
  Seljak, and Bertschinger}}]{Zaldarriaga:1997va}
\bibinfo{author}{\bibfnamefont{M.}~\bibnamefont{Zaldarriaga}},
  \bibinfo{author}{\bibfnamefont{U.}~\bibnamefont{Seljak}}, \bibnamefont{and}
  \bibinfo{author}{\bibfnamefont{E.}~\bibnamefont{Bertschinger}}
  (\bibinfo{year}{1997}), \eprint{astro-ph/9704265}.

\bibitem[{\citenamefont{Shankar}(1994)}]{Shankar:1994}
\bibinfo{author}{\bibfnamefont{R.}~\bibnamefont{Shankar}},
  \emph{\bibinfo{title}{Principles of Quantum Mechanics}}
  (\bibinfo{publisher}{Plenum Press}, \bibinfo{year}{1994}).

\bibitem[{\citenamefont{Csaki}(2004)}]{Csaki:2004ay}
\bibinfo{author}{\bibfnamefont{C.}~\bibnamefont{Csaki}} (\bibinfo{year}{2004}),
  \eprint{hep-ph/0404096}.

\bibitem[{\citenamefont{Cullen and Perelstein}(1999)}]{Cullen:1999hc}
\bibinfo{author}{\bibfnamefont{S.}~\bibnamefont{Cullen}} \bibnamefont{and}
  \bibinfo{author}{\bibfnamefont{M.}~\bibnamefont{Perelstein}},
  \bibinfo{journal}{Phys. Rev. Lett.} \textbf{\bibinfo{volume}{83}},
  \bibinfo{pages}{268} (\bibinfo{year}{1999}), \eprint{hep-ph/9903422}.

\bibitem[{\citenamefont{Sterman}(2004)}]{Sterman:2004pd}
\bibinfo{author}{\bibfnamefont{G.}~\bibnamefont{Sterman}}
  (\bibinfo{year}{2004}), \eprint{hep-ph/0412013}.

\bibitem[{\citenamefont{Pumplin et~al.}(2002)}]{Pumplin:2002vw}
\bibinfo{author}{\bibfnamefont{J.}~\bibnamefont{Pumplin}} \bibnamefont{et~al.},
  \bibinfo{journal}{JHEP} \textbf{\bibinfo{volume}{07}}, \bibinfo{pages}{012}
  (\bibinfo{year}{2002}), \eprint{hep-ph/0201195}.

\bibitem[{\citenamefont{{ATLAS collaboration}}(1999)}]{ATLAS:1999fr}
\bibinfo{author}{\bibnamefont{{ATLAS collaboration}}} (\bibinfo{year}{1999}),
  \bibinfo{note}{{CERN/LHCC/99/14-15}}.

\bibitem[{\citenamefont{Giudice et~al.}(2005)\citenamefont{Giudice, Plehn, and
  Strumia}}]{Giudice:2004mg}
\bibinfo{author}{\bibfnamefont{G.~F.} \bibnamefont{Giudice}},
  \bibinfo{author}{\bibfnamefont{T.}~\bibnamefont{Plehn}}, \bibnamefont{and}
  \bibinfo{author}{\bibfnamefont{A.}~\bibnamefont{Strumia}},
  \bibinfo{journal}{Nucl. Phys.} \textbf{\bibinfo{volume}{B706}},
  \bibinfo{pages}{455} (\bibinfo{year}{2005}), \eprint{hep-ph/0408320}.

\bibitem[{\citenamefont{Vacavant and Hinchliffe}(2001)}]{Vacavant:2001sd}
\bibinfo{author}{\bibfnamefont{L.}~\bibnamefont{Vacavant}} \bibnamefont{and}
  \bibinfo{author}{\bibfnamefont{I.}~\bibnamefont{Hinchliffe}},
  \bibinfo{journal}{J. Phys.} \textbf{\bibinfo{volume}{G27}},
  \bibinfo{pages}{1839} (\bibinfo{year}{2001}).

\bibitem[{\citenamefont{Sj{\"o}strand et~al.}(2006)\citenamefont{Sj{\"o}strand,
  Mrenna, and Skands}}]{Sjostrand:2006}
\bibinfo{author}{\bibfnamefont{T.}~\bibnamefont{Sj{\"o}strand}},
  \bibinfo{author}{\bibfnamefont{S.}~\bibnamefont{Mrenna}}, \bibnamefont{and}
  \bibinfo{author}{\bibfnamefont{P.}~\bibnamefont{Skands}},
  \bibinfo{journal}{JHEP} \textbf{\bibinfo{volume}{05}}, \bibinfo{pages}{026}
  (\bibinfo{year}{2006}), \eprint{0710.3820}.

\end{thebibliography}

\end{document}